\documentclass[twocolumn,showpacs,amsmath,nofootinbib]{revtex4-2}

\usepackage{float} 
\usepackage{amsmath}
\usepackage{amsfonts}
\usepackage{amssymb}
\usepackage{amsthm}
\usepackage{graphicx}
\usepackage[colorlinks,citecolor=blue]{hyperref}
\usepackage{color}
\usepackage{flafter}
\usepackage[normalem]{ulem}

\usepackage[toc,page]{appendix}
\usepackage{makecell}
\usepackage{slashed}
\usepackage{tikz-cd}
\usepackage{makecell}
\usepackage{subcaption}
\usepackage{graphicx}

\usepackage[font=small,labelfont=bf,
justification=justified,
format=plain]{caption}

\allowdisplaybreaks

\definecolor{OliveGreen}{rgb}{0,0.6,0}
\definecolor{Orange}{rgb}{1.00, 0.65, 0}
\definecolor{Grey}{rgb}{0.43, 0.5, 0.5}

\newcommand{\Fig}[1]{Fig.~\ref{#1}}
\newcommand{\Eq}[1]{Eq.(\ref{#1})}
\newcommand{\nn}{\nonumber\\}

\newcommand{\<}{\langle}
\renewcommand{\>}{\rangle}

\newcommand{\be}{\begin{eqnarray}}
\newcommand{\ee}{\end{eqnarray}}
\newcommand{\bpm}{\begin{pmatrix}}
\newcommand{\epm}{\end{pmatrix}}

\newcommand{\p}{\partial}

\newcommand{\Tr}{{\rm Tr}}

\newcommand{\ua}{\uparrow}

\newcommand{\da}{\downarrow}
\newcommand{\ra}{\rightarrow}

\renewcommand{\v}[1]{{\boldsymbol{#1}}}
\renewcommand{\a}{\alpha}
\renewcommand{\b}{\beta}

\newcommand{\s}{\sigma}
\renewcommand{\t}{\tau}

\newcommand{\G}{\Gamma}

\newcommand{\Avg}[1]{\langle #1 \rangle} 
\newcommand{\comment}[1]{}

\begin{document}

\title{Competing orders, the Wess-Zumino-Witten term, and spin liquids}
\author{Yen-Ta Huang$^{1}$}
\author{Dung-Hai Lee$^{1,2}$}\email{Corresponding author: dunghai@berkeley.edu}

\affiliation{
	$^1$ Materials Sciences Division, Lawrence Berkeley National Laboratory, Berkeley, CA 94720, USA.\\
	$^2$ Department of Physics, University of California, Berkeley, CA 94720, USA.\\
}

\begin{abstract}
In this paper, we demonstrate that in frustrated magnets when several conventional (i.e., symmetry-breaking) orders compete, and are ``intertwined'' by a Wess-Zumino-Witten (WZW) term, the possibility of spin liquid arises. The resulting spin liquid could have excitations which carry fractional spins and obey non-trivial self/mutual statistics.
As a concrete example, we consider the case where the competing orders are the N\'{e}el and valence-bond solid (VBS) order on square lattice. Examining different scenarios of  vortex condensation from the VBS side, we show that the intermediate phases, including spin liquids, between the N\'{e}el and VBS order always break certain symmetry. Remarkably, our starting theory, without fractionalized particles (partons) and guage field, predicts results agreeing with those derived from a parton theory. This suggests that the missing link between the Ginzberg-Landau-Wilson action of competing order and the physics of spin liquid is the WZW term. 
\end{abstract}
\maketitle

\newpage

\section{Introduction}

In the Landau paradigm, the phases of matter are characterized by broken symmetries.  In the last two decades, it is recognized that the phases of quantum material might not follow this paradigm. Consequently a frontier of condensed matter physics is the study of orders without broken symmetry. This includes symmetry-protected topological (SPT) order, such as topological insulator and superconductor \cite{Schnyder2008,Kitaev2009,Hasan2010}, and  “intrinsic topological order” such as that exhibited by spin liquids \cite{Anderson1987}. \\

Generally speaking, in 2+1 dimensions, theories of {\it non-chiral } spin liquids (for early examples, see Ref.\cite{Wen1991,Sachdev1991}) have two key ingredients.  (i) Spins fractionalize into partons. These partons usually couple to a gauge field, which tends to glue the partons back into the physical spins; (ii) A Higgs phenomenon, in which Higgs bosons carrying multiples of the fundamental gauge charge condense.  After (ii) the gauge field is (partially) gapped and the partons are deconfined \cite{Wen1991}. Moreover, the ungapped gauge flux has non-trivial mutual statistics with the particle carrying the fundamental gauge charge. In fact, (ii) can happen without (i). For example, the condensation of double vortices in a theory of U(1) bosons gaps the gauge field (which is dual to original boson phase), so that the gauge group becomes $\mathbb{Z}_2$.  The $\mathbb{Z}_2$ flux has mutual $-1$ Berry's phase with the single vortex. \\

In this paper we take a different perspective, where the starting point is a  Ginzburg-Landau-Wilson action describing the competition of {\it conventional} orders (hence no partons and no gauge field). However, the action contains a WZW term that ``intertwines'' different order parameters. 
In the concrete example we study, the competing orders are the N\'{e}el and valence-bond-solid (VBS) order of a frustrated magnet. The fractional spin is carried by the VBS vortex \cite{Levin2004,Tanaka2005,Senthil-Fisher2006}, the phase of the VBS order parameter acts as a gauge field, and spin liquid is triggered by the condensation of double vortices.\\     

The example described above is relevant to the frustrated spin 1/2 Heisenberg model. For the square lattice $J_1$-$J_2$ Heisenberg model, 
numerical evidence suggests the ground state evolves from N\'{e}el to VBS order as a function of increasing $J_2/J_1$. Interestingly, for a narrow interval of $J_2/J_1$ neither order is present.   It is widely felt that some sort of quantum spin liquid could be present in this $J_2/J_1$ interval.  This includes the critical $U(1)$ spin liquid associated with the ``deconfined quantum critical point'' \cite{Senthil2003}, as proposed in Ref.\cite{Gong2014}, and gapped \cite{Jiang2012} or gapless spin liquid \cite{Hu2013,Wang2013, Wang2018, Ferrari2020,Normura2021} phases.\\

This paper begins by deriving the theory of Ref.\cite{Tanaka2005}, namely, the VBS-N\'{e}el $O(5)$ non-linear sigma model with a WZW term (\Eq{o555}). We start with the famous parton theory (the $\pi$-flux mean-field theory) of Affleck and Marston \cite{Affleck1988}. Here the partons are massless Dirac fermions. These partons couple to a $SU(2)$ gauge field which, as we shall show, is confining. In Ref.\cite{Tanaka2005}, it is shown that a WZW term exists among the five mass terms of these massless fermions. However, this is done without worrying about the $SU(2)$ gauge field. In this paper, we begin by reviewing a derivation of the theory in Ref.\cite{Tanaka2005} in the presence of confining $SU(2)$ gauge field fluctuations \cite{Huang2021}.  \\

In a nutshell, the derivation amounts to the following. Without worrying about the $SU(2)$ gauge field, there are 36 possible mass terms, each corresponds to a parton-antiparton bound state. Borrowing the language of QCD, the partons are quarks and the mass terms are mesons.  
After the SU(2) confinement, only 5 mesons survive low energies. They correspond to the N\'{e}el and VBS order parameters in Ref.\cite{Tanaka2005}. Importantly, it is shown that the WZW term survives the confinement, leading to a five-component non-linear sigma model with a WZW term (\Eq{o555}).\\

The WZW term  predicts the  
the vortex of the VBS order possesses fractional spins ($S=1/2$) \cite{Levin2004, Tanaka2005, Senthil-Fisher2006}. We shall present a microscopic understanding of this spin fractionalization via the fermion zero modes, and show that it manifests the WZW term. Moreover, when the same WZW term is evaluated for the space-time process involving the exchange of vortices, it predicts the statistics of the vortex is bosonic. This is also confirmed by the fermion integration. Motivated by the double-vortex condensation in the U(1) boson theory, we consider different sequences of VBS  vortex condensation. Here we show the condensation of double VBS vortices triggers spin liquids. However, as we shall see, the spin liquids so obtained always break certain symmetry. \\

\begin{figure}
\centering
  \includegraphics[angle=0,scale=0.4]{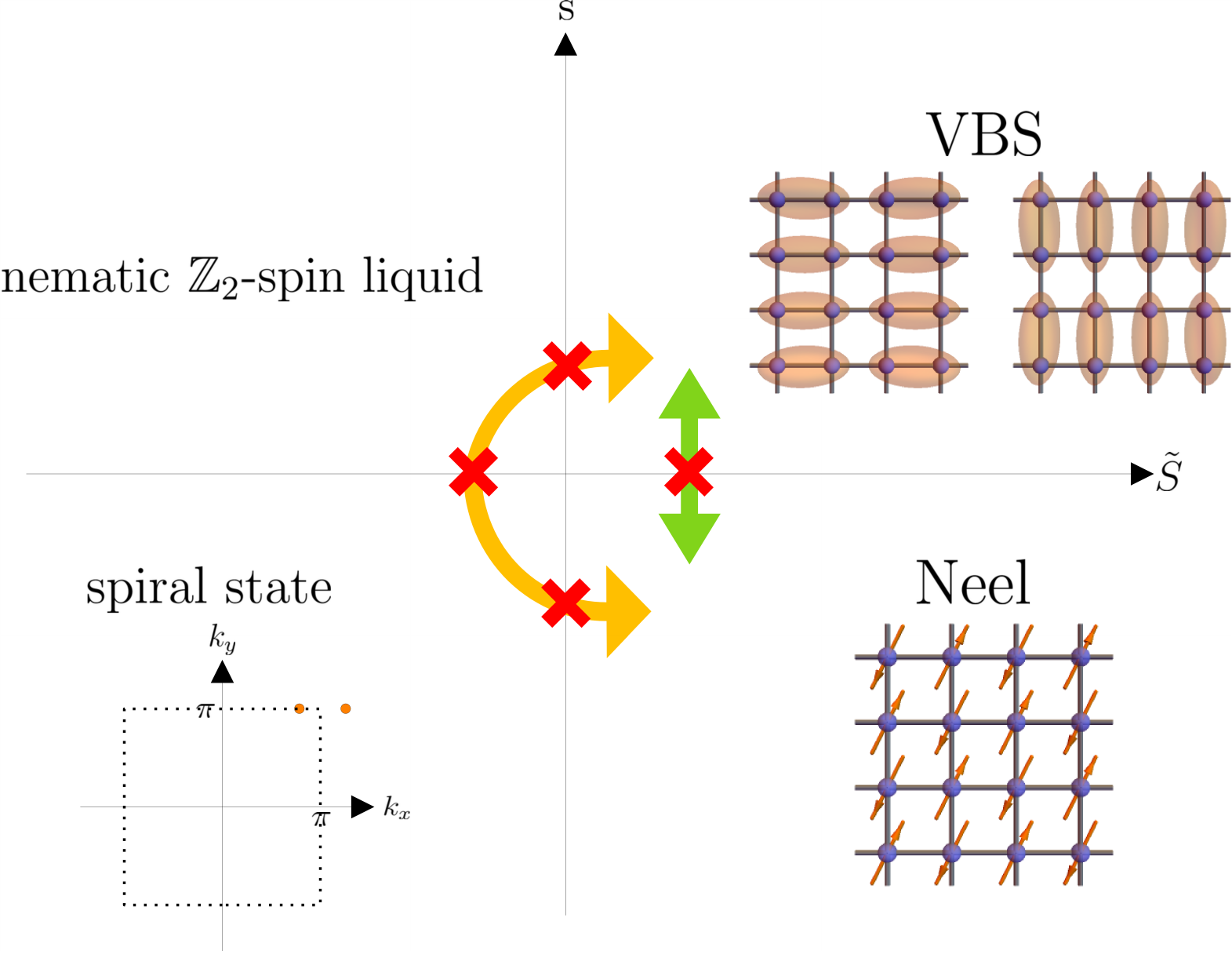}  
    \caption{The phase diagram where the green path realizes  the direct transition between the N\'{e}el and VBS phases. The red  cross marks the deconfined quantum critical point. The yellow path goes through a  nematic spin liquid and a uni-directional incommensurate spin density wave (SDW) (spin spiral state), phases  between the VBS and N\'{e}el states. The red crosses on the yellow path mark the respective quantum critical points. The inset in the lower-left corner shows the location of the SDW wavevector. Here $\tilde{S}$ is the mass of the spin-singlet double vortex and $s$ is the mass of spin 1/2 single vortex.
}
\label{NSL}
\end{figure}

\begin{figure}
  \centering
  \includegraphics[angle=0,scale=0.4]{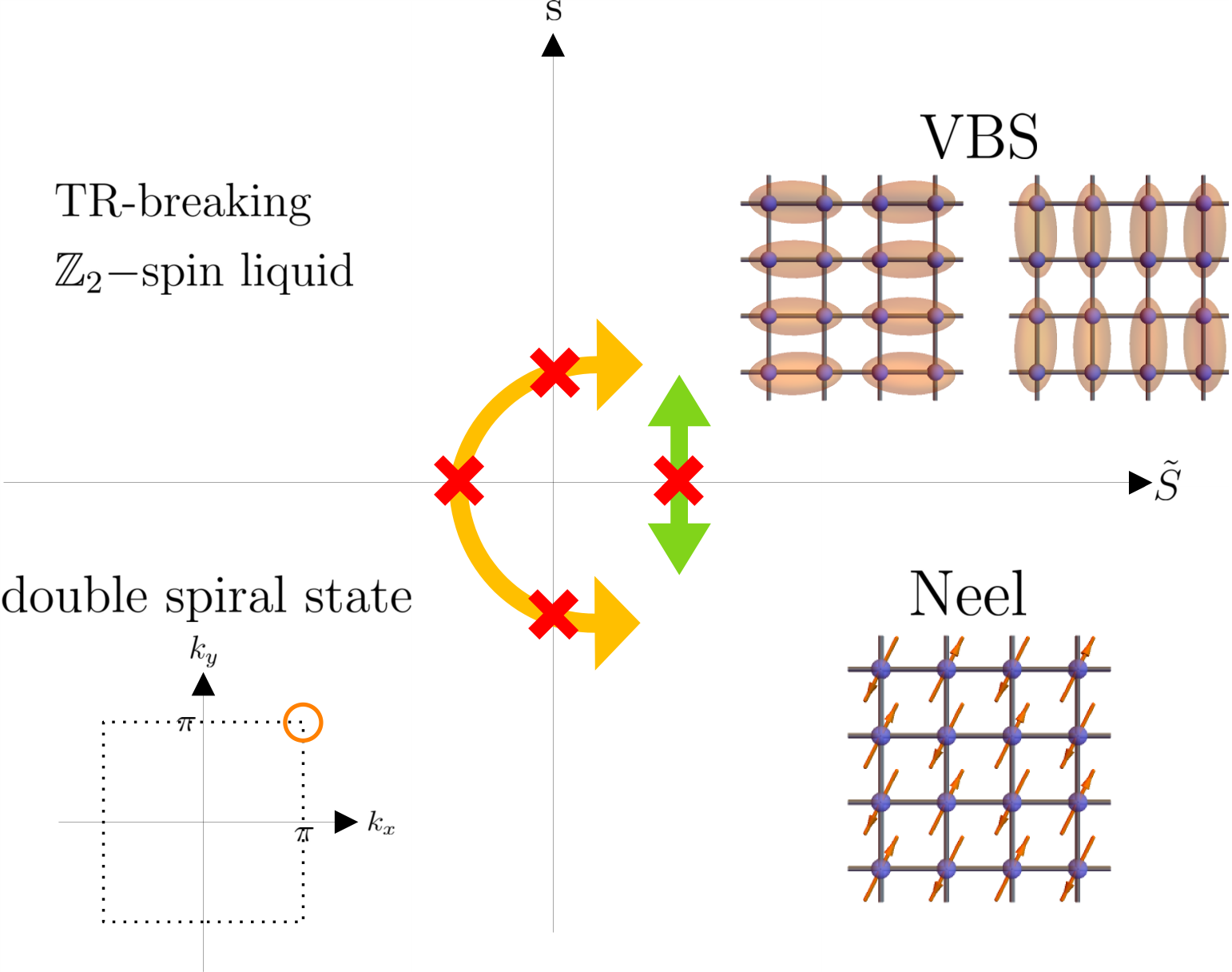}
		\caption{The phase diagram where the green path realizes  the direct transition between the N\'{e}el and VBS phases. The red cross marks the deconfined quantum critical point. The yellow path goes through a time-reversal breaking spin liquid and an isotropic double spiral  (SDW) phases  between the VBS and N\'{e}el states.  The red  crosses on the yellow path mark the respective quantum critical points. The inset in the lower-left corner shows the ring of ordering wavevector of the double spiral phase.  Here $\tilde{S}$ is the mass of the spin-singlet double vortex and $s$ is the mass of spin 1/2 single vortex.
}
	 \label{TSL}
\end{figure}

\begin{figure}
  \centering
		\includegraphics[scale=0.4]{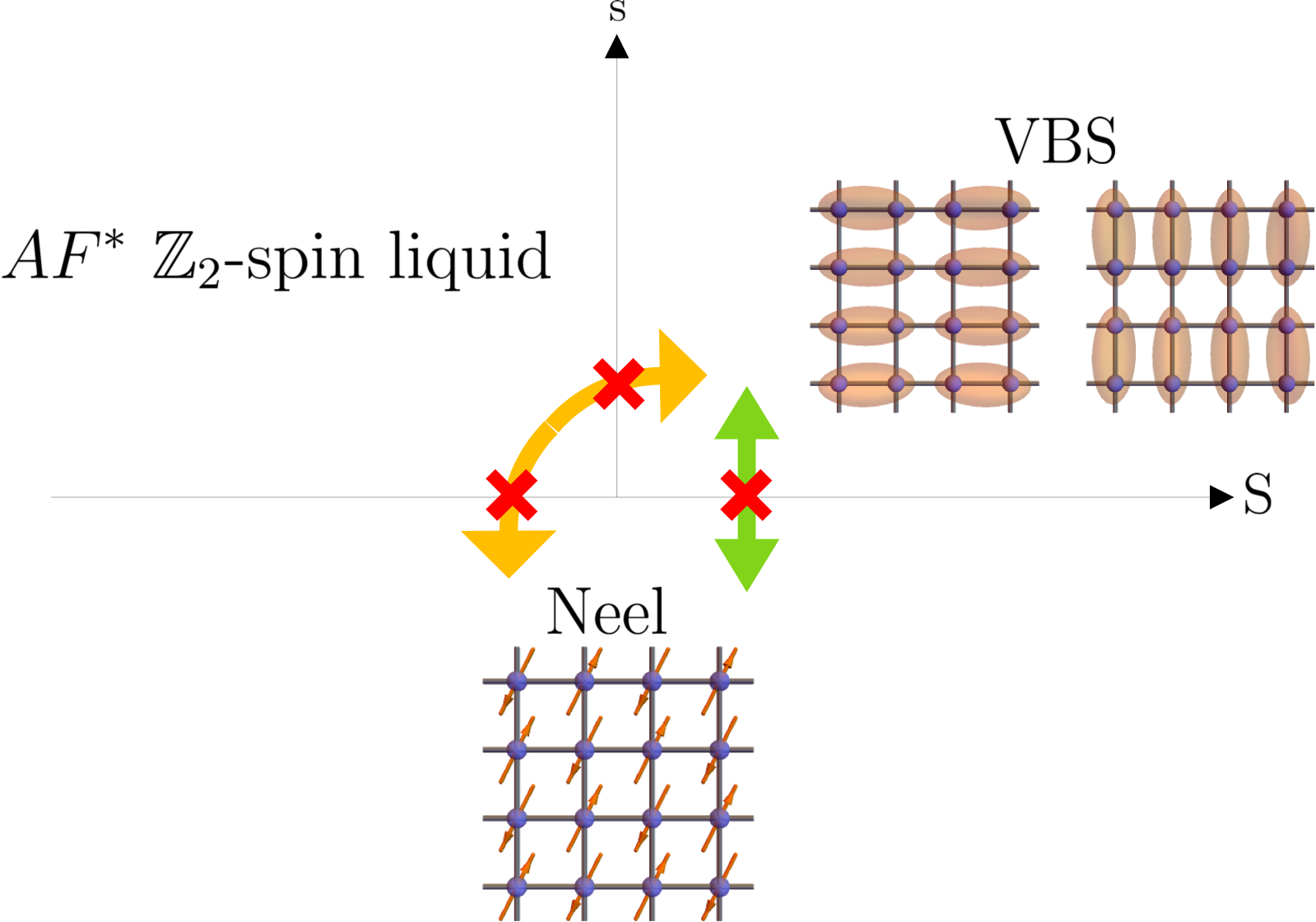}
		\caption{The phase diagram where the green path realizes  the direct transition between the N\'{e}el and VBS phases. The red  cross marks the deconfined quantum critical point. The yellow path goes through a topological ordered AF ordered phase  between the VBS and N\'{e}el states.  The red corsses on the yellow path mark the respective quantum critical points.  Here $S$ is the mass of the spin-triplet double vortex and $s$ is the mass of spin 1/2 single vortex.}
		\label{AFSL}
	\end{figure}

Our results are summarized by \Fig{NSL}, \Fig{TSL} and \Fig{AFSL}.  In each figure there are two alternative paths (marked as green and yellow) leading from VBS to N\'{e}el order. The green path realizes the direct transition between VBS and N\'{e}el order. It passes through the ``deconfined critical point'' (marked by the red cross) of Ref.\cite{Senthil2003}. The yellow paths, on the other hand, involve intermediate phases. These include different types of  $\mathbb{Z}_2$ spin liquids which break either the lattice, time reversal  or the spin rotation symmetry. They are accompanied by phases that break spin rotation symmetry. Here the red crosses mark the respective phase transitions. Remarkably, the scenario in \Fig{NSL} was predicted by the Schwinger-boson parton theory of Sachdev and Read \cite{Sachdev1991}. \\

It is important to point out that while the scenarios presented in \Fig{NSL} and \Fig{TSL} require the existence of further-neighbor, same-sublattice, spin interaction which favors spin singlet. In contrast scenario in \Fig{AFSL} requires the interaction to favor spin triplet. Therefore, \Fig{NSL} and \Fig{TSL} are more relevant for the frustrated Heisenberg model.  \\

The organization of this paper is as follows. In section \ref{pi_MF} we briefly review the derivation of the N\'{e}el-VBS non-linear sigma model with a WZW term. Here we start from partons which are 4-flavor Dirac fermions (it turns out that it is more convenient to view them as 8-flavor Majorana fermions)  representing the quasiparticles of the $\pi$-flux phase \cite{Affleck1988}. These fermions couple to a confining charge-$SU(2)$ gauge field \cite{Affleck1988b}. We will treat such confinement. In section \ref{core}, we derive the spin (with $S=1/2$) in the core of a single VBS vortex. Here we take two alternative approaches. In subsection \ref{swzw}, we determine the Berry phase of the vortex core from the WZW term, and identify it with the Berry phase of $S=1/2$ in 0+1 dimension. In subsection \ref{fermion_zero_mode}, we solve for the fermion zero modes in the vortex core, and show that the occupation of these zero modes leads to $S=1/2$ which is a charge-$SU(2)$ singlet, hence survives the confinement. The consistent results of subsections \ref{swzw} and \ref{fermion_zero_mode}  imply that theses two approaches are equivalent, i.e., the fermion zero mode is the manifestation of the WZW term. 
In section \ref{stwzw} we determine the statistics of the VBS vortex by computing  the Berry phase associated with exchanging two vortices.  In \ref{exwzw} we compute this phase using the WZW term, and in \ref{stpart} we compute it by fermion integration. Again, the consistency of the results in subsections \ref{exwzw} and \ref{stpart} implies the equivalence of the two approaches. In section \ref{trsv} we determine the projective representation of one-lattice-constant  (of the square lattice with the original $1\time 1$ unit cell) translation carried by the fermionic partons,  and use it to derive the transformation law of  $\hat{\Omega}$. The same transformation law can be deduced by inspecting \Fig{NeelVBS}.  Using this transformation law, we deduce the effects of translation on the VBS vortex, and construct the vortex field theory in Section \ref{vft}. In section \ref{VBSNeel}, we study the condensation of the single VBS vortex which leads to the direct VBS-N\'{e}el transition shown as the green paths in \Fig{NSL} to \Fig{AFSL}.  In section \ref{cssdv}, we study the scenario in which the condensation of spin-singlet double VBS vortices precedes the condensation of single vortices. Section \ref{cssdv} contains 4 subsections. In subsection \ref{NNSL} and \ref{icsdw}, the double vortex is a $p$-wave pair of two single vortices, while in subsection \ref{TRBSL} and \ref{dspsdw} the double vortex is a $p_x+ip_y$ pair of two single vortices. The discussions in these sections lead to \Fig{NSL} and \Fig{TSL}. 
In section \ref{AF*}, we study the scenario where the spin-triplet double VBS vortex condenses. This leads to \Fig{AFSL}.	As the concluding discussion, section \ref{condis} includes the relation of the current work with the parton approaches, and the effects of hole doping. In addition to the main text, there is one appendix \ref{boson_int} which presents details of integrating out the massive single vortices when the double vortices have condensed.\\

\section{The $O(5)$ non-linear sigma model ($\Eq{o555}$) derived in a nutshell}
\label{pi_MF}\hfill

In this section we review the  theory presented in Ref.\cite{Huang2021}. At half-filling, the low energy degrees of freedom of the Hubbard model are  spins. The fermion parton representation amounts to representing the spin operator as  bilinears of fermion (``spinon'') operators, namely, 
\be
S_i^a =\frac{1}{2} f_{i\a}^\dagger \sigma^a_{\a\b} f_{i\b}.
\label{stspn}
\ee
The Hilbert space of spins is recovered after imposing the single occupation constraints \be &&f_{i\ua}^\dagger f_{i\ua}+f_{i\da}^\dagger f_{i\da}=1\nn
&&f^\dagger_{i\ua}f^\dagger_{i\da}=0\nn&&f_{i\da}f_{i\ua}=0.
\label{constraint}\ee
Because the last two lines of \Eq{constraint} break the spinon number conservation, it's convenient to rewrite the spinon operators using Majorana fermions 
\begin{align*}
f_{i \alpha} := F_{i ,1 \alpha} + i F_{i, 2 \alpha}.
\end{align*}
In terms of the Majorana operators, the spin operators are represented as
\be
&&S_i^a =\frac{1}{2} F_i^\dagger \Sigma^a F_i,~~{\rm where}\nn
&&\Sigma^a = \left(-YX, IY, -YZ \right),
\label{sop1}
\ee
where $F_i$ has four components. In the last line of \Eq{sop1}, I,X,Y,Z stand for Pauli matrices $\s_0,\s_x,\s_y,\s_z$, and two Pauli matrices standing next to each other denotes tensor product. In \Eq{sop1}, the first and second Pauli matrices carry  the Majorana and spin indices respectively. Using  $F_i$ the occupation constraints read
\be
&&  F^T_i \left( XY \right) F_i :=F^T_i T^1 F_i=0 \nn
&& F^T_i \left(ZY \right) F_i :=F^T_i T^2 F_i=0 \nn
&& F^T_i \left(-YI \right) F_i :=F^T_i T^3 F_i=0.
\label{constr22}\ee
\\

In the Mott insulating phase, the low energy Hamiltonian is a function of spin operators. Expressed in terms of the parton (Majorana fermion) operators, even the simplest bilinear spin-spin interaction involves four fermion operators. By (i) introducing the auxiliary fields $\chi_{ij}$ and $\Delta_{ij}$ that decouple such four fermion operators into fermion bilinear, and (ii) impose the constraint in \Eq{constr22} by space-time dependent Lagrange multipliers $a^1_{i0}, a^2_{i0}, a^3_{i0}$, we obtain
the following spinon path integral \cite{Lee2006} $$
Z=\int D[F] D[\chi]D[\Delta]D[a_0] \exp{\left(-S\right)} 
$$
where
\be
S&&=\int_0^\b d\t \Big\{\sum_i F^T_{i}\p_0 F_{i} + \sum_{\Avg{ij}} \frac{3}{8} J_{ij} \Big[ F_i^T\Big(Re[\chi_{ij} ] YI\nn&&+ i Im[\chi_{ij}] II  + Re[\Delta_{ij}] XY - Im[\Delta_{ij}] ZY \Big) F_j\nn
&&+|\chi_{ij}|^2 + |\eta_{ij}|^2 \Big] - i\sum_i   a^b_{i0}  \left( F^T_i T^b F_i\right)\Big\}.
\label{spap1}
\ee
In \Eq{spap1} $J_{ij}$ is the exchange constant between spins on site $i$ and $j$.
\\

\subsection{The $\pi$-flux saddle point}\hfill

\begin{figure}
	\centering
		\includegraphics[scale=0.32]{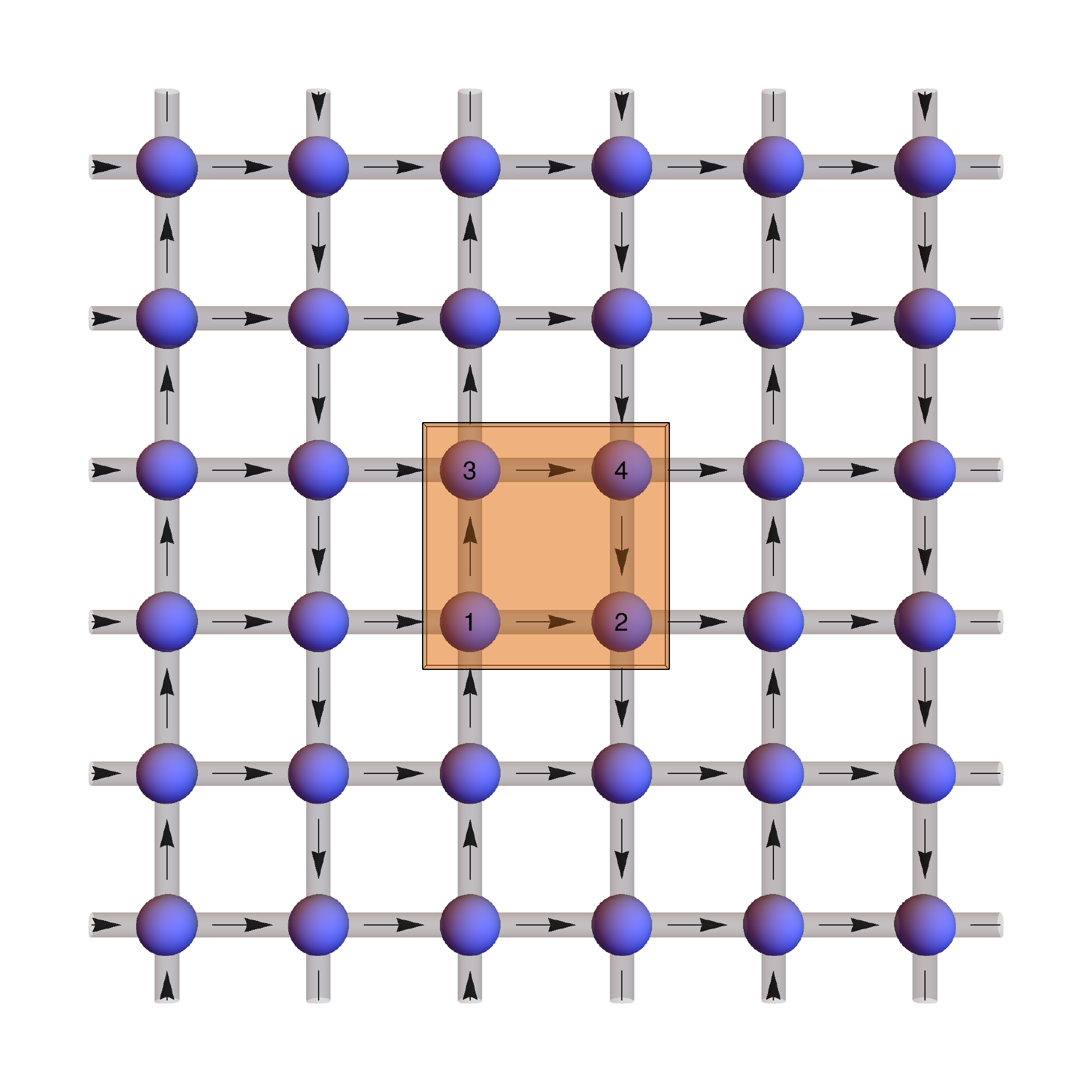}
		\caption{The hopping term in the $\pi$-flux phase saddle point solution of \Eq{spap1}. The orange square labels the 4-site unit cell.}
		\label{pi_Flux1}
\end{figure}

The Affleck-Marston  $\pi$-flux phase is a saddle point solution of \Eq{spap1} where $\{ J_{ij} \}$ only extends to the nearest neighbors\footnote{In the following we shall start with the nearest-neighbor interaction and derive \Eq{o555}. The further neighbor interactions will be incorporated later on, in the non-linear sigma model as the ... in \Eq{o55}. This approach is useful so long as the N\'{e}el and VBS remain to be the most prominent competing order parameters.}. It corresponds to 
\be
	\bar{\Delta}_{ij} = 0 , ~~ \bar{\chi}_{i+\hat{x},i} = i\chi , ~~ \bar{\chi}_{i+\hat{y},i} = i (-1)^{i_x} \chi, ~~ \bar{a}_{i0}^{1,2,3} = 0.\nn
	\label{pimf}
\ee
\noindent where $\chi$ is a  real number.
The hopping pattern in \Eq{pimf} is shown in \Fig{pi_Flux1}, where the arrows point in the direction of $+i\chi$ hopping.

Choosing a 4-site unit cell and define $\psi^T=(F_1,F_2,F_3,F_4)^T$,  the momentum-space mean-field Hamiltonian read
\be
\hat{H}_{\rm MF} =
 - \frac{3}{8}J \chi \, \sum_{\bf k} \psi_{\bf-k}^T \, \left[ II \otimes 
h(\v k)\right] \psi_{\bf k}, 
\label{mjmf}
\ee
where
\be
&&h(\v k)=\nn&&\begin{bmatrix}
0&-i(1-e^{ik_1})&-i(1-e^{ik_2})&0\\
i(1-e^{-ik_1})&0&0&i(1-e^{ik_2})\\i(1-e^{-ik_2})&0&0&-i(1-e^{ik_1})\\0&-i(1-e^{-ik_2})&i(1-e^{-ik_1})&0
\end{bmatrix}\nonumber
\ee

The fermion dispersion is given by
\begin{align*}
	E(\v k)&=\pm {3J\over 8}\chi\sqrt{4-2 \left( \cos k_1 + \cos k_2 \right)},
\end{align*}
where each branch has 8-fold degeneracy.

The fermion low energy modes occur near the node ${\v k}_0=0.$ Writing $\v k = {\v k}_0 + {\v q}$ and expanding the mean-field Hamiltonian  to linear order in $\v q$ 
we obtain 
\be
\hat{H}_{\rm MF} =	\frac{3J}{4} \chi \sum_{\v q}\psi^T(- \v q) (q_1 \Gamma_1 + q_2 \Gamma_2 ) \psi(\v q),
	\label{dirac}\ee
	where 
\be 	\Gamma_1 = IIIX, ~ \Gamma_2 =IIXZ.\label{gma}\ee

In \Eq{gma} the first two Pauli matrices  correspond to the Majorana and spin degrees of freedom. The indices of the tensor product of the last two Pauli matrices label the $2\times 2$ unit cell in  \Fig{pi_Flux1}, namely, 
\begin{align*}
	ZI=+1,~IZ=+1:& ~ \text{site 1} \\
	ZI=+1,~IZ=-1:& ~ \text{site 2} \\
	ZI=-1,~IZ=+1:& ~ \text{site 3} \\
	ZI=-1,~IZ=-1:& ~ \text{site 4}
\end{align*}
Fourier transforming the Hamiltonian back to the real space, we obtain
\be \hat{H}_{\rm MF} =\frac{3J}{16 } \chi ~\int d^2r~ \psi^T(\v r)\left(-i\Gamma_j \partial_j\right) \psi(\v r), \label{MFH}\ee  where 
$\psi(\v r)=\sum_{\v q} \psi(\v q) e^{i\v q\cdot \v r}.$ \Eq{MFH} is the point of departure for the following discussions.

\subsection{The $SU(2)$ gauge field and the confinement}\hfill 

Because the operators in \Eq{sop1} are invariant under the following local ``charge-SU(2)''
transformation
\be
\label{su2g}
&F_j\ra e^{i\theta_{b,j}t^b} F_j,~t^b =( XY, ZY,-YI)~{\rm or}\nn
&\psi(\v r)\ra e^{i\theta_{b}(\v r)T^b} \psi(\v r),~T^b =( XYII, ZYII,-YIII),\nn
\ee
we expect that the spinon field theory should involve a  charge-$SU(2)$ gauge field. We stress that this $SU(2)$ is different from the spin $SU(2)$, which is generated by $$\Sigma^a = \left(-YXII, IYII, -YZII \right).$$ 
The spatial components of the charge $SU(2)$ gauge field  are identified as   $$U_{ij}=\bar{U}_{ij}e^{i a_{ij}},$$ where $a_{ij}=a_{ij}^b T^b$, while $a_{i0}^b$ act as the time component of the charge-$SU(2)$ gauge field \cite{Lee2006}. Thus, the continuum version of the low energy theory is given by 
\be
	\label{Eq:gauge_matter}
	S &&= \int d\t d^2r \Big\{\psi^T \left[ (\partial_0 - i a_0 ) - i \frac{3J}{16 } \chi ~\Gamma_i (\partial_i - i a_i) \right] \psi \nn&&+ {1\over 2g} \Tr[f_{\mu\nu}f_{\mu\nu}]
	\Big\}
\ee
\noindent where $\G_i$ is given by \Eq{gma}, and $a_\mu =a_{\mu, b} T^b$ with $T^b$ given by \Eq{su2g}. The last Maxwell term in \Eq{Eq:gauge_matter} is generated by integrating out the high energy fermion degrees of freedom.\\

\begin{figure}
	\begin{subfigure}{0.2\textwidth}
		\includegraphics[scale=0.25]{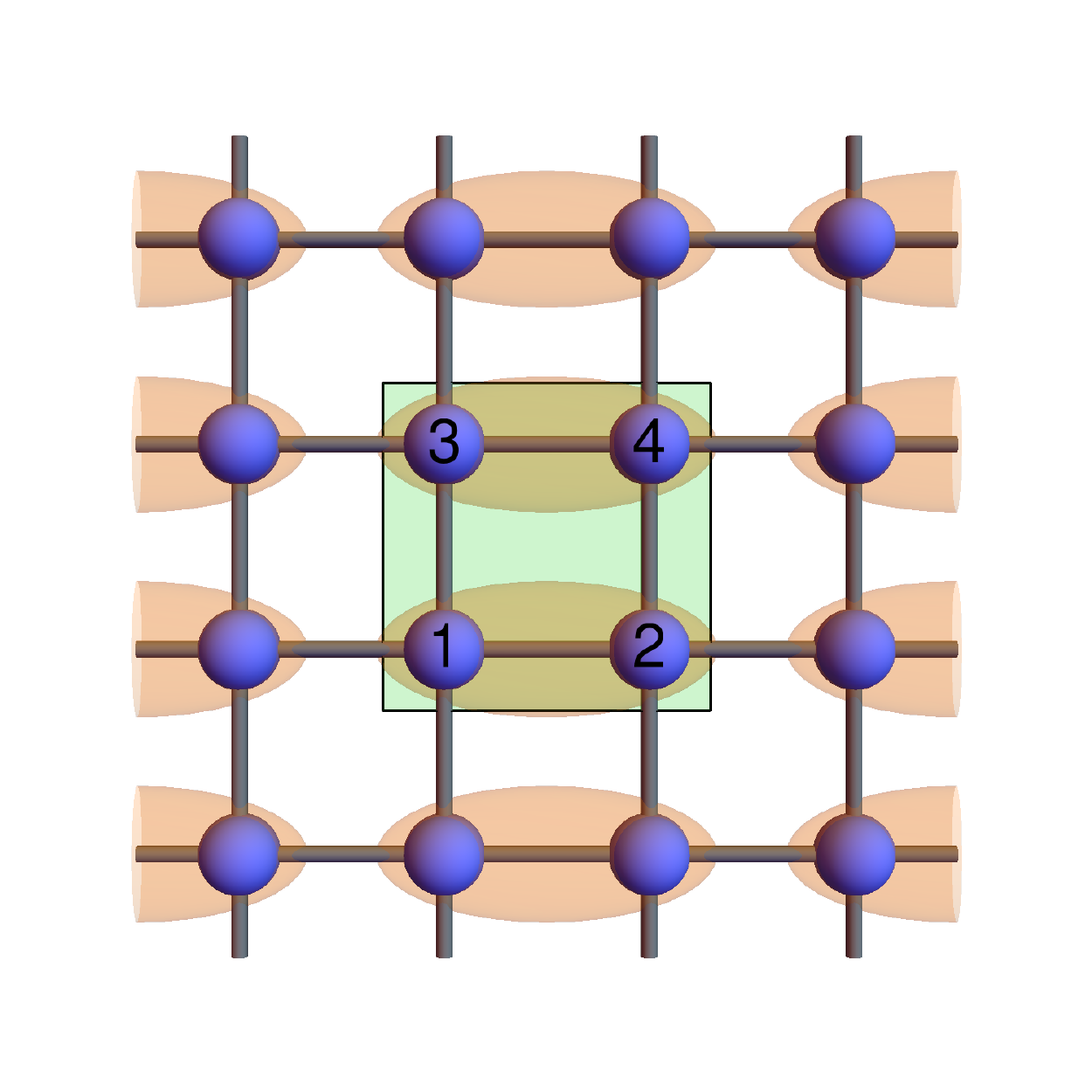}
		\captionsetup{font=normalsize,labelfont=normalsize}
		\caption{} 
	\end{subfigure}
\hspace*{0cm}
	\begin{subfigure}{0.2\textwidth}
		\includegraphics[scale=0.25]{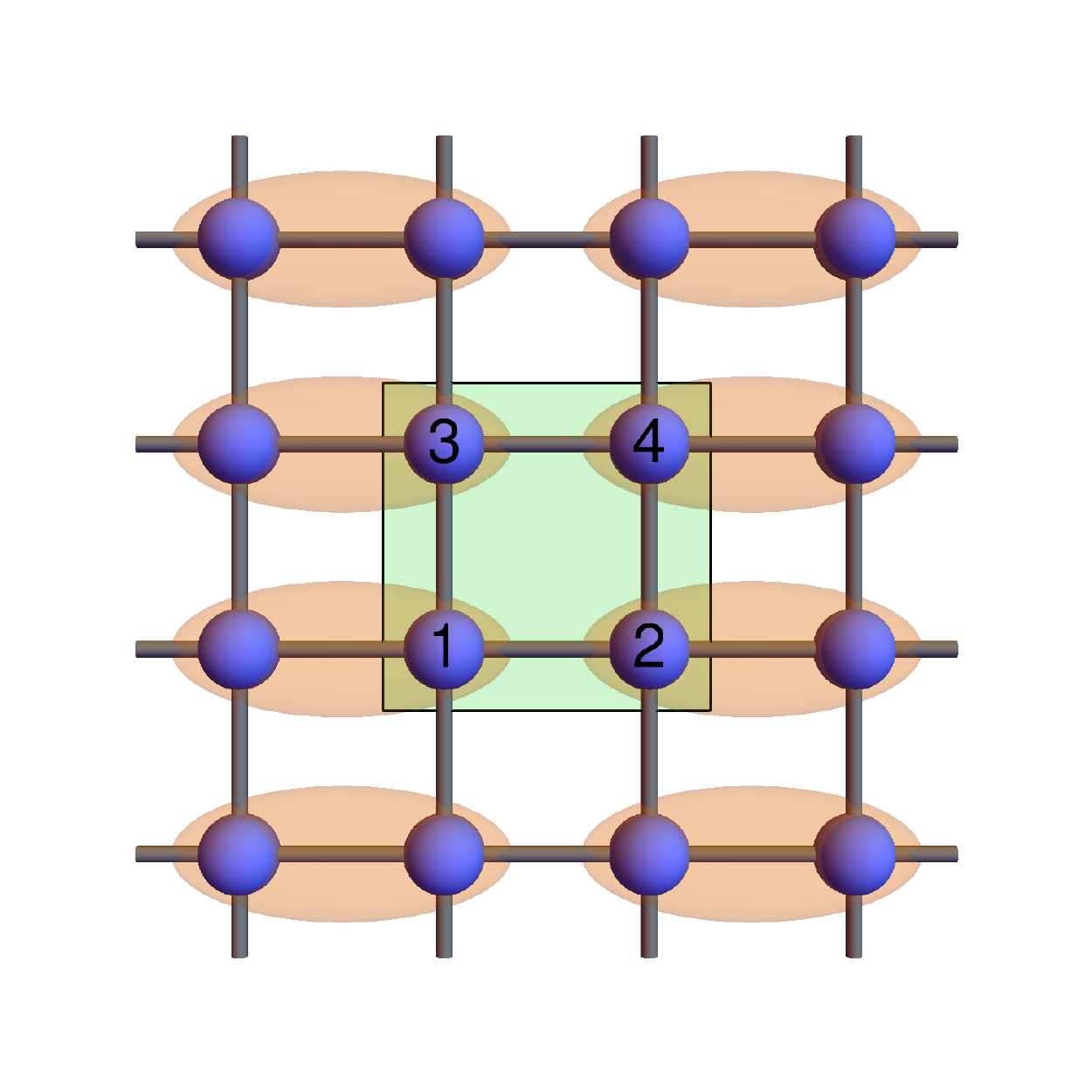}
		\captionsetup{font=normalsize,labelfont=normalsize}
		\caption{} 
	\end{subfigure}
\vspace*{.cm}   
	\begin{subfigure}{0.2\textwidth}
		\includegraphics[scale=0.25]{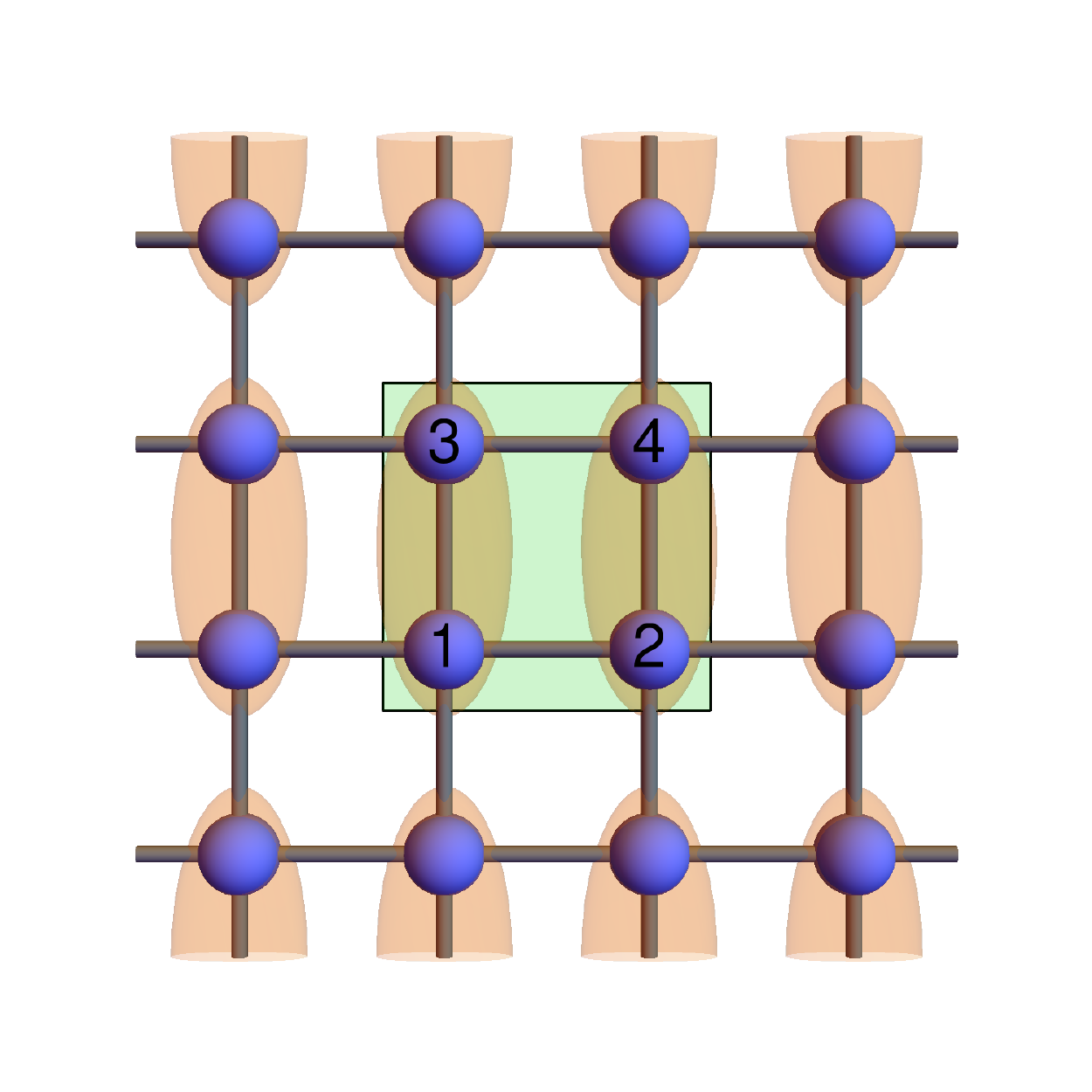}
		\captionsetup{font=normalsize,labelfont=normalsize}
		\caption{} 
	\end{subfigure}
	\begin{subfigure}{0.2\textwidth}
		\includegraphics[scale=0.25]{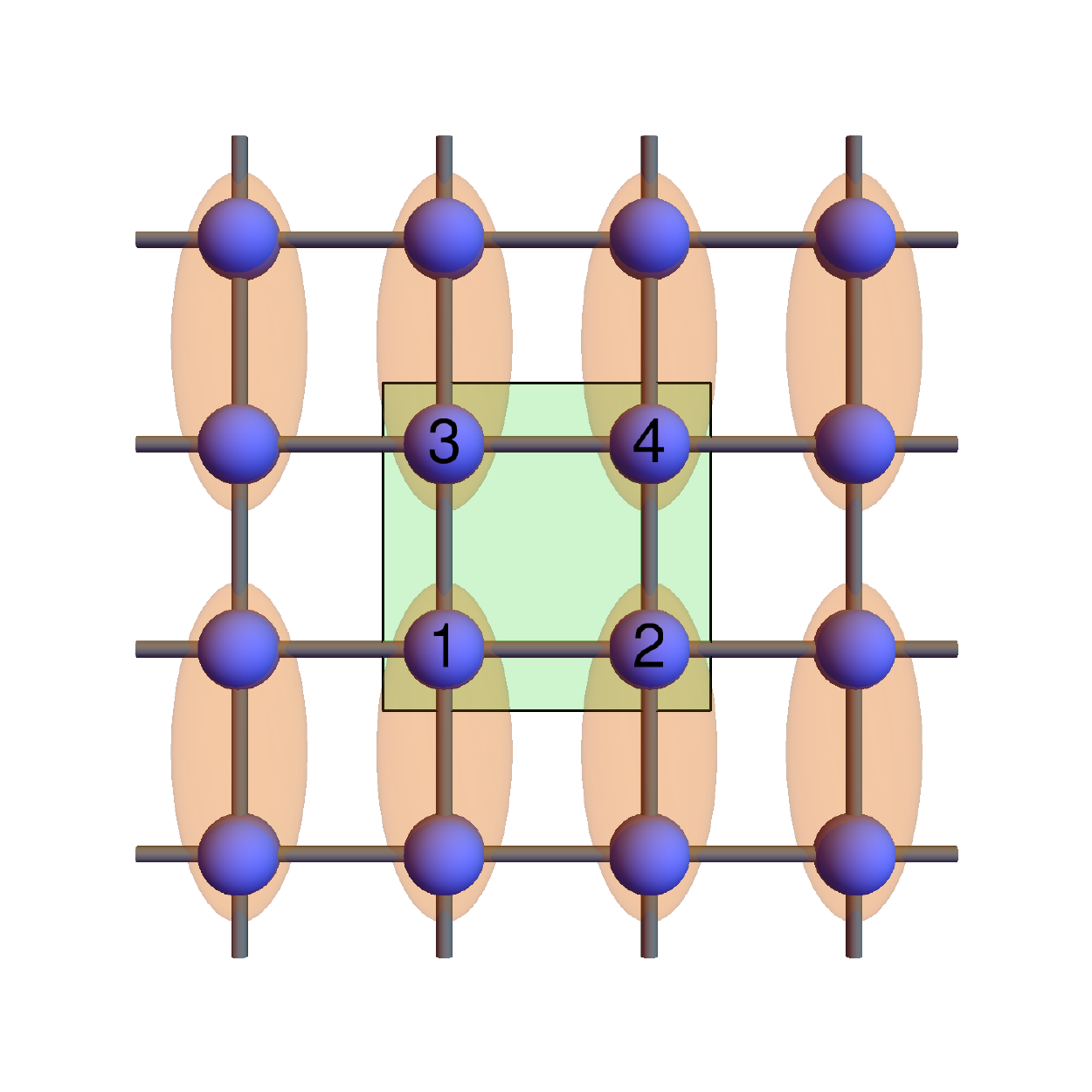}
		\captionsetup{font=normalsize,labelfont=normalsize}
		\caption{} 
	\end{subfigure}
\vspace*{.cm}  
	\begin{subfigure}{0.2\textwidth}
		\includegraphics[scale=0.25]{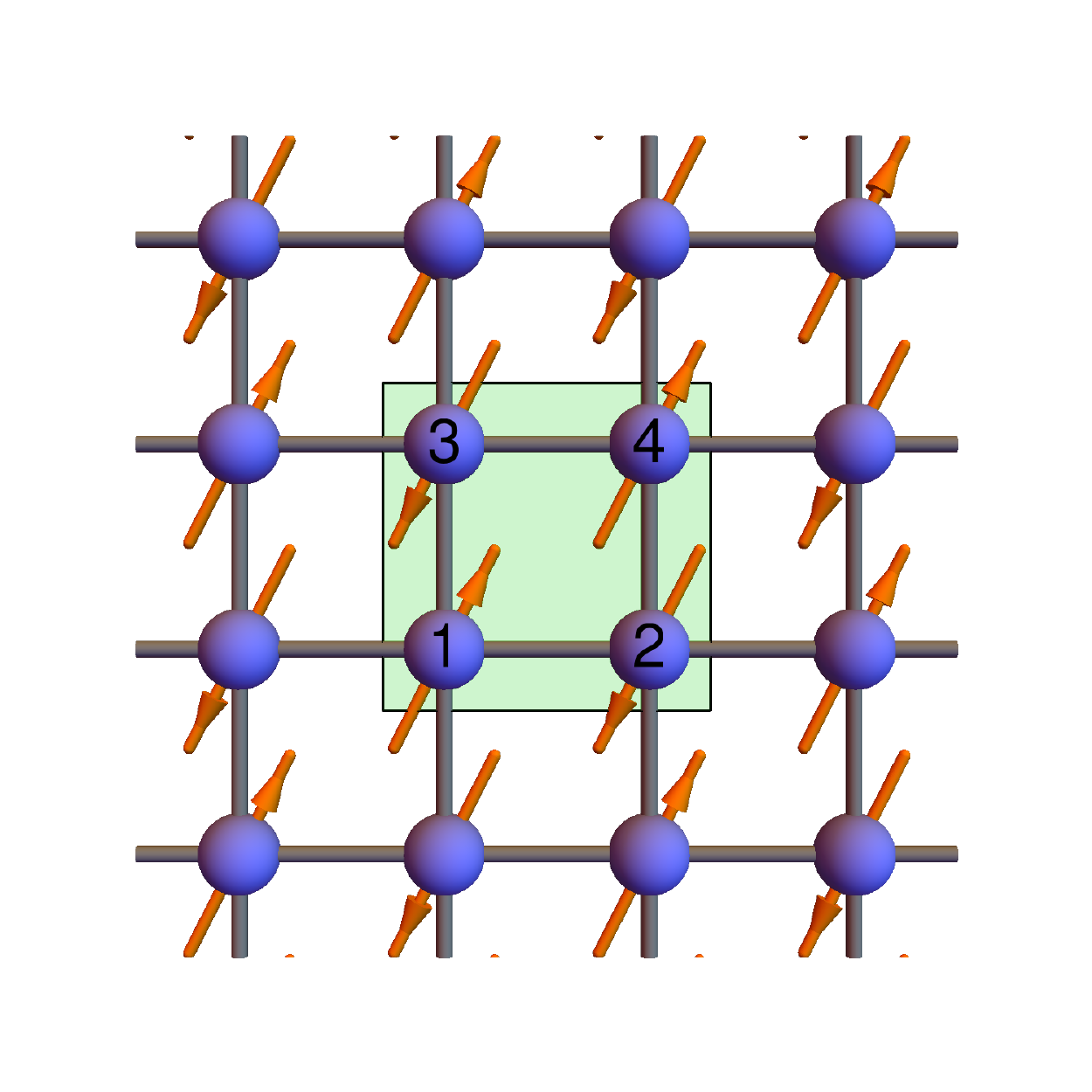}
		\captionsetup{font=normalsize,labelfont=normalsize}
		\caption{} 
	\end{subfigure}
\caption{Spin configurations associated with the VBS and N\'{e}el order. The green box marks a unit cell. The orange ellipse denotes spin-singlets. The first four figures represent VBS order with (a) $v_1>0, v_2=0$, (b) $v_1<0, v_2=0$, (c) $v_1=0, v_2>0$, (d) $v_1=0, v_2<0$. Figure (e) is the N\'{e}el order in specific $(n_1,n_2,n_3)$ direction.
	}
	\label{NeelVBS}
\end{figure}

Via non-abelian bosonization \cite{Huang2021}, it is shown in the absence the SU(2) gauge field, there are 35 mass (gap-opening) terms for \Eq{dirac} relevant to the discussion here. Each of these mass terms has the form \be\psi^TM_i\psi\label{masss}\ee and corresponds to a spinon-antispinon bilinear.  Borrowing the language of QCD, the order parameters correspond to these mass terms are the meson fields. Under the constraint that the total gap is a constant, the meson fields   form the manifold ${O(8)\over O(4)\times O(4)}$, and the ${O(8)\over O(4)\times O(4)}$ non-linear sigma model with a WZW term governs the dynamics of the mesons. After integrating out the charge-SU(2) gauge field, the spinons are confined into five $SU(2)$ singlet mesons with the $M_i$ in \Eq{masss} given by
\be
M_i = YXZZ, IYZZ, YZZZ, IIIY, IIYZ.\label{meson}\ee These mesons form an  $S^4$ sub-manifold in ${O(8)\over O(4)\times O(4)}$.
Substituting these meson fields into the ${O(8)\over O(4)\times O(4)}$ non-linear sigma model, we obtain \Eq{o555},
\begin{align}
	W[\hat{\Omega}] =&{1\over 2g}\int\limits_{\mathcal{M}} d^3 x \, \left( \partial_\mu \Omega_i \right)^2+W_{\rm WZW}[\hat{\Omega}]. \label{o555}
\end{align}
Here the order parameter $\hat{\Omega}=(n_1,n_2,n_3,v_1,v_2)$ is a five-component unit vector associated with the masses $(M_1,M_2,M_3,M_4,M_5)$ in \Eq{meson}. Because $M_{1,2,3}$ are the masses that correspond to the N\'{e}el order, and $M_{4,5}$ are to the VBS order, (see \Fig{NeelVBS}), 
we identify \Eq{o555} as the non-linear sigma model governing the N\'{e}el-VBS competition.   Due to the charge-$SU(2)$-singlet nature of the masses, \Eq{o555} is immune to the $SU(2)$ gauge fluctuations!

We stress that aside from the WZW term, \Eq{o555} is the ``fixed-length'' version of the Ginzberg-Landau-Wilson action of five competing order parameters. $W_{\rm WZW}$, given by \Eq{wzw10}, is pure imaginary with quantum mechanical origin. As we shall see, it is the missing link between classical competing order and spin liquid physics. 

\subsection{The Wess-Zumino-Witten term}\hfill

The WZW term is a Berry's phase, namely, 
\be
W_{WZW}[\tilde{\hat{\Omega}}] =- \frac{2\pi i }{64 \pi^2} \int\limits_{\mathcal{B}} \epsilon^{ijklm} \tilde{\Omega}_i \, d\tilde{\Omega}_j  \, d\tilde{\Omega}_k \, d\tilde{\Omega}_l \, d\tilde{\Omega}_m. \nn
\label{wzw10}\ee
In \Eq{wzw10} $\tilde{\hat{\Omega}}(\tau,x,y,u)$ represents a one-parameter-family extension of the space-time configuration $\hat{\Omega}(x,y,\t)$. At $u=0$, the configuration is trivial, say, $\tilde{\hat{\Omega}}(\tau,x,y,0)=(0,0,0,0,1)$, At $u=1$ the  $\hat{\Omega}$ realizes the physical configuration. It can be shown that $\exp\left(-W_{WZW}\right)$ is independent of the specific choice of the one-parameter family as long as the coefficient in \Eq{wzw10} is an integer multiple of  $\frac{2\pi i }{64 \pi^2}$.   
\footnote{In order for the interpolation between the trivial and non-trivial configurations to exist, it requires any configuration of the order parameters can be smooth deformed into a trivial configuration. This requires the homotopy group of the space-time to the order parameter space map to be trivial $$\pi_{3}(S^4)=0.$$  To ensure $\exp\left(-W_{WZW}\right)$ to be independent of the specific path of interpolation it requires  $$\pi_{4}(S^4)=\mathbb{Z}.$$ Of course both of these requirements are satisfied by the five component order parameter space $S^4$.} 
\section{The spin of a VBS vortex}
\label{core}

\subsection{The WZW term restricted to the vortex core}\label{swzw}\hfill

In this subsection, we show that the WZW term in \Eq{wzw10} predicts that the core of a VBS vortex harbors a spin with $S=1/2$.\\ 

\begin{figure}
	\centering
	\includegraphics[scale=0.4]{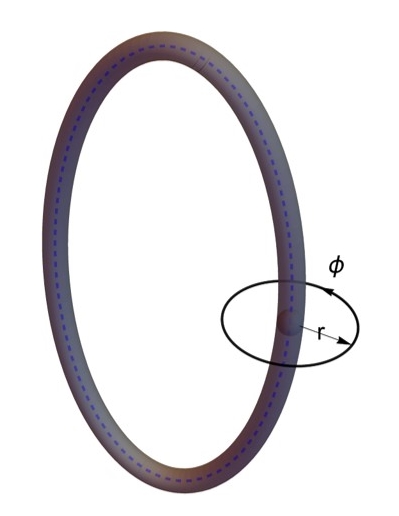}
	\caption{The tube of radius $r_c$ around a vortex loop (the blue dashed line).}
	\label{verloop}
\end{figure}

In a VBS vortex configuration, $\hat{\Omega}=(0,0,0,v_1,v_2)$ with $v_1^2+v_2^2=1$ everywhere except close to the vortex core, 
where  $v_1^2+v_2^2\ra 0 $ while $n_1^2+n_2^2+n_3^2\ra 1$. Let's compute the Berry phase associated with a single vortex loop, which is a one-dimensional closed curve parametrized by $\tau_v$ (the blue dashed line in \Fig{verloop}). The core region, drawn as the gray tube in \Fig{verloop}, has radius $r_c$. Thus the tube region forms $S^1 \times D^2$ where $S^1$ is the vortex loop and $D^2$ is the 2D disk with radius $r_c$. Within the tube, $(n_1,n_2,n_3)$  has non-zero magnitude. 

Let's  use $r$ and $\phi$ to parametrize the local radial distance from the vortex loop and the angle around it (see \Fig{verloop}). The configuration of $\hat{\Omega}$ is given by
\be
\hat{\Omega}(r,\phi,\tau_v)=\left(\sqrt{1-f^2(r)}\hat{n}(\tau_v),f(r) \cos \phi, f(r)\sin \phi\right)\nn 
\label{vloop}\ee
where $f(r)$ is a smooth function satisfying $f(r) \ra 1$ for  $r>r_c$, and $f(r) \rightarrow 0 $ as $r\ra 0$.
Since the vortex core is microscopic in size, in \Eq{vloop} we have assumed that there is no {\it spatial} (i.e., $r,\phi$) variation of $\hat{n}$. Nonetheless, the direction of $\hat{n}$ can vary along the vortex loop (i.e. depends on $\tau_v$). 
To evaluate the WZW term, we smoothly deform \Eq{vloop} to a configuration where $\hat{n}$ is constant along the vortex loop (due to $\pi_1(S^2)=0$ such a deformation exists). Let $u \in [0,1]$ parametrizes the deformation, i.e., 
$\tilde{\hat{n}}(\tau_v,u=0) = \hat{n}(\tau_v)$ while $\tilde{\hat{n}}(\tau_v,u=1) = \hat{z}$.
The corresponding deformed $\hat{\Omega}$ is then given by
\be
\tilde{\hat{\Omega}}(r,\phi,\tau_v,u)=(\sqrt{1-f^2(r)} ~ \tilde{\hat{n}}(\tau_v,u),  f(r) \cos \phi, f(r) \sin \phi).\nn
\label{vloop2}\ee
Pluging \Eq{vloop2} into \Eq{wzw10}, we obtain,
\be
W_{\rm{WZW}}
= -2\pi i \left(\frac{1 }{4\pi}\int\limits_{D^2 } du \, d\tau_v \,    \epsilon^{abc} \tilde{\hat{n}}_a \, \partial_u \tilde{\hat{n}}_b  \, \partial_{\tau_v}\tilde{\hat{n}}_c \right).   \nn
\label{corebp}\ee
\noindent \Eq{corebp} is the Berry phase associated with a $S=1/2$ spin in the $(0+1)$-dimensional coherent state path integral. This shows  there is $S=1/2$ spin residing in the core of the VBS vortex.  

\subsection{The fermion zero modes}
\label{fermion_zero_mode}\hfill

In the following, we study the spinon zero modes in the VBS vortex core. The spinon Hamiltonian associated with a vortex read
\be
&&\hat{H}_v=\int d^2r~\psi^T(\v r) h_v(\v r) \psi(\v r)~{\rm where~}\nn
&&h_v(\v r)=\left[ -i \G_1 \partial_1 - i \G_2 \partial_2 + v_1(\v r) M_4 + v_2(\v r) M_5 \right].
\label{zmh}
\ee
Here 
$\G_1,\G_2$ are given by \Eq{gma} and $M_4,M_5$ are given by \Eq{meson}. 
To map the zero mode solution of \Eq{zmh} to a solved problem, namely, Ref.\cite{Jackiw1981}, we interchange the order of lower-right and upper-left sites in  the unit cell (i.e.,  $2 \leftrightarrow 4$ in \Fig{pi_Flux1}). 
This is achieved by the following 
orthogonal transformation $II\otimes R$ where
\be
	R =  \begin{pmatrix} 
			1 & 0 & 0 & 0 \\
			0 & 0 & 0 & 1 \\
			0 & 0 & 1 & 0 \\
			0 & 1 & 0 & 0						 
					\end{pmatrix}.
\label{sitere}\ee
Under this change of basis \be &&h_v(\v r)\ra  (II \otimes R) \cdot h_v(\v r) \cdot (II\otimes R)^\dagger\nn
&&= II\otimes \begin{pmatrix} 0 & \mathcal{Q}(\v r) \\ \mathcal{Q}^\dagger(\v r) & 0 \end{pmatrix},
\ee
where \be \mathcal{Q}= -i \left[X\partial_1 + Z\partial_2 + v_1(\v r) E + v_2(\v r)I\right].\label{Qeq}\ee
The fermion zero modes are the normalizable zero energy eigenvectors satisfying 
\be
&&\mathcal{Q}^\dagger \begin{pmatrix} u_1 \\ u_2\end{pmatrix}=0, u_3=u_4=0~~{\rm or}\nn	 &&\mathcal{Q} \begin{pmatrix} u_3 \\ u_4\end{pmatrix}=0,u_1=u_2=0.
\label{zm00}\ee
Given  $v_1(\v r)$ and $v_2(\v r)$ forming a vortex pattern, \Eq{Qeq} and \Eq{zm00} were solved by Jackiw and Rossi \cite{Jackiw1981}. 
\\

Because the sites in each unit cell have been reordered, namely, 
\be &&1\leftrightarrow{\rm~ lower-left~},2\leftrightarrow{\rm~upper-right~},\nn&&3\leftrightarrow{\rm~upper-left~},4\leftrightarrow{\rm~lower-right~}.\label{relab}\ee
\noindent  the zero mode localizes on the lower-left  and upper-right lattice sites if there is a normalizable solution for 
\be
		\mathcal{Q}^\dagger \begin{pmatrix} u_1 \\ u_2\end{pmatrix}= 0.
		\label{zm11}\ee
		Similarly, it is localized on  the upper-left and lower-right lattice sites if there is a normalizable solution for
\be
	\mathcal{Q} \begin{pmatrix} u_3 \\ u_4\end{pmatrix}=0.
	\label{zm22}
\ee
Ref.\cite{Jackiw1981} explicitly solved the differential equation. Their solution shows that, depending on the vorticity, the zero mode is localized on different sublattices.  Specifically when the vorticity is +1, the zero mode localizes on site 1 and 4, and when the vorticity is $-1$, the zero mode localizes on site 2 and 3. Note that (1,2) and (3,4) belong to different sublattices of the square lattice. Finally, each Majorana zero mode has a degeneracy 4, which corresponds to two complex fermion zero modes -- one for each spin.

After unit cell site reordering in \Eq{relab}, $IIZI=\pm 1$ corresponds to the two sublattices of the square lattice, and the zero mode of vorticity  $\pm 1$ has $IIZI=\pm 1$. Within each of the $IIZI=\pm 1$ subspace, the $IIIZ=\pm 1$ components of the zero mode are related by \Eq{zm11} or \Eq{zm22}. In other words, for a vortex of fixed vorticity ($\pm 1$), each Majorana zero mode is 4-fold degenerate. 
It is straightforward to show that the projection of 
$T^b$ in \Eq{su2g} onto the degenerate subspace is given by 
\begin{align*}
\tilde{a}_\mu = \sum_{k=1}^3\tilde{a}_\mu^b t_b, ~ \text{ where } t_b= \frac{1}{4}(XY,ZY,-YI).
\end{align*} 
\noindent This is the same as the charge-$SU(2)$ generator for a single site. Thus, the charge-SU(2) singlet constraint requires that only one of the spin zero modes can be occupied. 
Finally, turning on $\hat{n}$ in the vortex core splits the zero modes and  produces a $S=1/2$ magnetic moment.  \\

The above analysis has implications later on, when we consider hole doping. Namely, hole doping removes the core spins of the VBS vortex. This is consistent with the effects of hole doping in the t-J model.

\section{The statistics of the VBS vortex}\label{stwzw}
\subsection{The WZW term restricted to the vortex exchange process}\label{exwzw}\hfill

We can determine the statistics of the VBS vortex by computing the Berry phase associated with the space-time configuration $\hat{\Omega}$ in which two vortices are exchanged. Technically, because our space-time is $S^3$, we need to consider the process in which   two pairs of vortex-anti-vortex are created out of vacuum, exchange the two vortices, and then annihilate the vortices with the anti-vortices at the end. Such Berry's phase contains two contributions: 1) the Berry phase due to exchange of the vortices and 2) the Berry phase associated with the spin 1/2  in the vortex/antivortex cores. To isolate the Berry phase due to the exchange, we lock the core spins in, say, the positive $n_3$-direction.

In the following, we present an argument which suggests that under the space-time configuration discussed above, the Berry phase arising from the WZW term in \Eq{wzw10} is zero.  To argue that the Berry phase is zero involves two steps. (i) Using the result in Ref.\cite{Abanov2000}, one can show that when any {\it one} of the five components of $\hat{\Omega}$ is zero, the WZW term reduces to the topological $\theta$-term  (with $\theta=\pi$) for the remaining four components. This topological term is non-zero only when the wrapping number associated with the mapping from the space-time to the order parameter manifold, in this case $S^3$, is non-zero. (ii) We note that in the present situation only three components of $\hat{\Omega}$ are non-zero. By counting the dimension of the space-time image, we conclude that any such $\hat{\Omega}$ cannot produce a non-zero wrapping number in $S^3$, hence the Berry phase vanishes.%

\subsection{Fermion integration}\label{stpart}\hfill

The $\hat{\Omega}$ configuration discussed in the preceding subsection represents a map from the space-time ($S^3$) to the order parameter space $S^2$ formed by $(n_3,v_1,v_2)$. In general, such a mapping allows for a Hopf term.\\

Let's go back to the Majorana fermion representation. Integrating out the fermion gives rise to the WZW term in \Eq{wzw10}. The difference now is that only   $(n_3,v_1,v_2)$  are non-zero and the corresponding spinon masses are given by $(M_3,M_4,M_5)$ in \Eq{meson}.  
Because both the gamma matrices in \Eq{gma} and all the mass terms in  \Eq{meson}  conserve the fermion number, (the associated $U(1)$ symmetry is generated by $Q=YIII$), 
it's convenient to complexify the Majorana fermion so that
\begin{align*}
	\Gamma_1 \ra& IXX, ~ \Gamma_2 \ra IXZ \\
	M_3 \ra& ZZI, ~ M_4\ra IXY,~M_5\ra IYI.
\end{align*}
These matrices commute with $ZII$, and hence can be block-diagonalized into the $ZII=\pm 1$ sectors. Within these each sectors, the fermion integration leads to a coefficient $\pm \pi$ Hopf term (see \cite{Abanov2000,Huang2021}). Thus, the net Hopf terms cancel, resulting in a vanishing Berry's phase.
\\

Therefore both subsection \ref{exwzw} and subsection \ref{stpart}  points to the fact that the $S=1/2$ VBS vortex is a boson. 

\section{Translating the vortex}
\label{trsv}
\hfill

Due to the hopping phase in \Fig{pi_Flux1}, translation by one lattice constant (of the  square lattice with the original $1\times 1$ unit cell) is realized projectively. This is because in order to restore the mean-field Hamiltonian, translation must be accompanied by a gauge transformation. Restoring the original site labeling in \Fig{pi_Flux1}, it can be shown that the combined transformations are given by
\begin{align*}
	&\begin{pmatrix}\psi(\v r)_{\alpha \sigma 1} \\ \psi(\v r)_{\alpha \sigma 2} \\\psi(\v r)_{\alpha \sigma 3} \\ \psi(\v r)_{\alpha \sigma 4} \end{pmatrix}  \xrightarrow{\hat{T}_{\hat{\v x}}} \begin{pmatrix}\psi(\v r)_{\alpha \sigma 2} \\ \psi(\v r +  \hat{ \v x})_{\alpha \sigma 1} \\ - \psi(\v r)_{\alpha \sigma 4} \\ -\psi(\v r + \hat{\v x})_{\alpha \sigma 3} \end{pmatrix} , \\
	 ~~&\begin{pmatrix}\psi(\v r)_{\alpha \sigma 1} \\ \psi(\v r)_{\alpha \sigma 2} \\\psi(\v r)_{\alpha \sigma 3} \\ \psi(\v r)_{\alpha \sigma 4} \end{pmatrix}  \xrightarrow{\hat{T}_{\hat{\v y}}} \begin{pmatrix}\psi(\v r)_{\alpha \sigma 3} \\ \psi(\v r)_{\alpha \sigma 4} \\ - \psi(\v r + \hat{\v y})_{\alpha \sigma 1} \\ -\psi(\v r + \hat{\v y})_{\alpha \sigma 2} \end{pmatrix}
\end{align*}
where the subscripts of $\psi$ are $\alpha=$ Majorana index,  $\sigma=$ the spin index, and $1,2,3,4$ are the site indices in the unit cell (\Fig{pi_Flux1}).
\noindent This gives the following transformation matrices for the low energy spinon modes,
\begin{align*}
	&T^{\hat{\v x}} = IIZX , ~ T^{\hat{\v y}} = IIXI.
	\end{align*}
As before, the Pauli matrices are ordered according to 
	$$\text{Majorana}\otimes\text{spin}\otimes 4\times 4~\text{unit-cell sites}$$
When acted on the masses of \Eq{meson}, this leads to the following transformations of the VBS and N\'{e}el order parameters 
\begin{align*}
	(n_1, n_2, n_3, v_1, v_2) \xrightarrow{\hat{T}_x}& (-n_1, -n_2, -n_3, -v_1, v_2) \\
	(n_1, n_2, n_3, v_1, v_2) \xrightarrow{\hat{T}_y}& (-n_1, -n_2, -n_3, v_1, -v_2) 
\end{align*}
The same conclusion can be reached without worrying about the fermion projective transformations by directly inspecting \Fig{NeelVBS}.
\\

Because reversing the sign of $v_1$ or $v_2$ changes the vorticity of a VBS vortex, and due to the discussion in section \ref{fermion_zero_mode}, we conclude that translation results in particle-antiparticle conjugation on the spin label of (the relativistic)  $\phi_\s$, namely,
\be
	\phi_\sigma \xrightarrow{\hat{T}_x, \hat{T}_y} E_{\sigma \sigma'} \phi_{\sigma'}^*.
\label{prjt}\ee
This transformation law will be useful later when we discuss the symmetries of the vortex field theory.\\

We note that many of the results in this section, e.g., vortex spin, the effect of translation on vortex field, are readily obtained by Levin and Senthil\cite{Levin2004} using  picturesque arguments. 

\section{The field theory of single VBS vortices}
\label{vft}
\hfill

\Eq{o555}  possesses an emergent symmetry  (the $O(5)$ symmetry) which is not shared by the microscopic model. Therefore generically we should supplement \Eq{o555} with additional $O(5)$ symmetry breaking terms, i.e,  
\begin{align}
	W[\hat{\Omega}] =~&{1\over 2g}\int\limits_{\mathcal{M}} d^3 x \, \left( \partial_\mu \Omega_i \right)^2+W_{\rm WZW}[\hat{\Omega}]\nn+~&O(5)~\text{symmetry breaking terms} \label{o55}
\end{align}
These additional terms can be generated by, e.g.,  integrating out the high energy fermion modes, or by integrating out the charge-$SU(2)$ gauge fluctuations. A simple example of such term is
$
\mathcal{D} \int d^3 x \, \left(\sum_{i=1}^3 n_i^2 - \sum_{i=1}^2 v_i^2 \right),
$
which tips the balance between the N\'{e}el and VBS order. 
\\

In the rest of the paper we shall start from the VBS phase and look at the possible vortex proliferations that can eliminate the VBS order \cite{Levin2004}.

As seen in the  previous section, the VBS vortex is a  spin-$1/2$ boson\footnote{Here we emphasize that the spin is a flavor degree of freedom.  It is not the ``spin'' in the Dirac sense, otherwise, the spin-statistics theorem would be violated.}.  In addition, the core-spin resides on opposite sublattice for vortex and anti-vortex. Hence conservation of vorticity implies that the vortex hops among the same sublattice. In terms of the vortex annihilation operator $\phi_\sigma$, the action for a relativistic boson field $\phi_\s$ has the form 
\begin{align}
\label{singlev}
S_{\phi} =& \int d^3x \Big[ |(\partial_\mu - i a_\mu ) \phi_\s|^2  + s |\phi_\s |^2 + \frac{v}{2} |\phi_\s|^4  \\
+& \frac{1}{2g} (\epsilon_{\mu\nu\rho} \partial_\nu a_\rho )^2 \Big], \notag
\end{align}
	where $\s=\ua,\da$, and $a_\mu$ is the $U(1)$ gauge field (not to be confused with the charge-$SU(2)$ gauge field, which had been integrated out) whose fluctuation is the dual representation of the phase fluctuation of the VBS order parameter
	$\psi_{\rm VBS}=v_1+i v_2.$
	The theory in \Eq{singlev} can be obtained by  the usual $U(1)$ boson duality of the VBS order parameter $\psi_{\rm VBS}$, except that the vortex core is decorated with $S=1/2$. 

\Eq{singlev} has the following symmetries,
\be
&&\text{Vorticity conservation} : \phi_\s \rightarrow  e^{i\eta} \phi_\s, \nn
&&\text{$U(1)$ gauge symmetry} : \phi_\s \rightarrow  e^{i\tilde{\eta}(x)} \phi_\s,~~a_\mu\ra a_\mu+\p_\mu\tilde{\eta}(x) \nn
&&\text{Flavor (spin) $SU(2)$:} \phi_\s \rightarrow  u_{\s\s'} \phi_{\s'} \text{, where } u_{\s\s'} \in SU(2) \nn
&&\text{$T_{\hat{x}}$, $T_{\hat{y}}$: } \phi_\s \rightarrow E_{\s\s'} \phi_{\s'}^* \nn
&&\text{$T$: } \phi_\s \rightarrow E_{\s\s'} \phi_{\s'}
\label{sym}
\ee
Note that unlike ordinary translations, $T_x^2=T_y^2=-1$.
	\\
	
As it stands the gauge field in \Eq{singlev} is non-compact. However, because of the underlying square lattice, the VBS order parameter is subjected to a 4-fold anisotropy\footnote{The preferred angles of $\psi_{\rm VBS}$ will depend on whether we have the columnar or plaquette VBS order.}. This means that $a_\mu$ is in fact a compact gauge field, and in carrying out the path integral we need to include the space-time events where degree 4 monopole in $a_\mu$ can pop in/out.  In the phase where such monopole condenses, the vortices are confined and the VBS long-range order prevails. However, we shall be interested in condensing various types of VBS vortices in the following. In those phases $a_\mu$ is necessarily deconfining. Therefore to address the properties of these vortex condensed phases,  we shall ignore the monopoles and start with    
\Eq{singlev}.

\section{The condensation of single vortices -- the direct VBS to N\'{e}el transition}\label{VBSNeel}
\hfill

In Ref.\cite{Levin2004} it is proposed that the condensation of $\phi_\s$ leads to the N\'{e}el order. Moreover, it is argued that the de-confined quantum critical point (DCQCP) between VBS and N\'{e}el order is realized when the $s$ in \Eq{singlev} is tuned from positive to negative so that $\<\phi_\s\>\ne 0$.
Such single vortex condensation leads to  the N\'{e}el state with the antiferromagnetic order parameter given by
\begin{align*}
	\vec{n} =\phi_\alpha^* \vec{\sigma}_{\alpha\beta} \phi_\beta .
\end{align*}
This direct VBS to N\'{e}el transition is marked by the red cross on the green paths in \Fig{NSL}-\Fig{AFSL}.  
\\

In the next three sections we shall study the scenarios in which preceding the single vortex condensation, the double vortex (with vorticity $\pm 2$) condenses first. Since a double vortex is the bound state of two single vortices, and the single vortex is a spin 1/2 boson,  exchange symmetry implies  the double vortex can be either 1) spin-singlet and odd-parity, or 2) spin-triplet and even-parity. The fact that spin-singlet double vortex has to be odd in parity is responsible for the breaking of the point group or time reversal symmetry  in the coming sections. \\

Note that such broken symmetry depends on the geometry of the underlying lattice. In Ref.\cite{Fa2010} it is pointed out that on the honeycomb lattice it is possible for spin-singlet odd parity double vortex to preserve the  point group symmetry. 

\section{The condensation of spin-singlet odd-parity double vortex}\label{cssdv}
\hfill

In terms of $\phi_\s$, the double vortex annihilation operator is given by
\be \tilde{\Phi}_\a(\v x)=\frac{1}{\sqrt{2}} \phi^T_a(\v x) E_{ab} \partial_\alpha \phi_b(\v x). \label{dbls}\ee
Here $\a=1,2$ label the spatial directions, $a,b$ labels the spins, and $E_{ab}$ is the two-index anti-symmetric tensor. Recall that on bipartite lattice, vortices with the same vorticity resides on the same sublattice, hence for the square lattice $\a=1$ means ${x+y\over 2}$ and $\a=2$ means ${x-y\over 2}$. 
\\

The action which involves both the single and double vortex fields is given by
\be
	\label{Sphi-Phit}
	&&S_{\phi,\tilde{\Phi}} = \int d^3 x \Big\{ \frac{1}{2g} (\epsilon_{\mu\nu\rho} \partial_\nu a_\rho )^2+|(\partial_\mu - i a_\mu ) \phi_\sigma|^2\nn&& +|(\partial_\mu - i 2 a_\mu ) \tilde{\Phi}_\alpha|^2+V_{\phi_\s}[\phi_\s]+V_{\tilde{\Phi}_\alpha}[\tilde{\Phi}_\alpha]\nn
	&&-\lambda \left[ \tilde{\Phi}_\alpha^* \left(\frac{1}{\sqrt{2}} \phi^T E \partial_\alpha \phi \right) + \text{c.c.}~\right]
	\Big\}.
\ee
In \Eq{Sphi-Phit}
\be
&&V_{\phi_\s}[\phi_\s]= s |\phi_\sigma |^2 + \frac{v}{2} |\phi_\sigma|^4\nn
&&V_{\tilde{\Phi}_\alpha}[\tilde{\Phi}_\alpha]= \tilde{S} |\tilde{\Phi}_\alpha |^2 
	+ \frac{V_1}{4} (|\tilde{\Phi}_1|^4 + |\tilde{\Phi}_2|^4) + V_2 |\tilde{\Phi}_1|^2 |\tilde{\Phi}_2|^2\nn
	&&+ \frac{W}{2} \left( (\tilde{\Phi}_1^* \tilde{\Phi}_2)^2 + \text{c.c.} \right),
\label{pote}\ee
and the $\lambda$ term  enforces \Eq{dbls}.
\Eq{Sphi-Phit} is consistent with the result of Ref.\cite{Sachdev1991,Thomson2018}. 
\\

\noindent  {\it On square lattice}, the quartic terms in $V_{\tilde{\Phi}_\alpha}[\tilde{\Phi}_\alpha]$ are constructed to be invariant under the spatial $90$ degree rotation, inversion, and time-reversal transformations
\begin{align}
	\begin{pmatrix} \tilde{\Phi}_1 \\ \tilde{\Phi}_2 \end{pmatrix} & \xrightarrow{\text{$90$ deg rotation}} \begin{pmatrix} \tilde{\Phi}_2 \\ -\tilde{\Phi}_1 \end{pmatrix} \notag\\
	\begin{pmatrix} \tilde{\Phi}_1 \\ \tilde{\Phi}_2 \end{pmatrix} & \xrightarrow{\text{Inversion}} \begin{pmatrix} -\tilde{\Phi}_1 \\ -\tilde{\Phi}_2 \end{pmatrix} \notag\\
	\begin{pmatrix} \tilde{\Phi}_1 \\ \tilde{\Phi}_2 \end{pmatrix} & \xrightarrow{\text{~time-reversal~~}}  \begin{pmatrix} \tilde{\Phi}_1 \\ \tilde{\Phi}_2 \end{pmatrix}. 
	\label{symm_Phi}
\end{align}
\\

Let's first focus on $V_{\tilde{\Phi}_\alpha}[\tilde{\Phi}_\alpha]$ . Assume $$\tilde{\Phi}_\alpha = \rho_\alpha e^{i \Theta_\alpha},$$ the potential energy is given by
\begin{align*}
	V_{\tilde{\Phi}_\alpha}[\tilde{\Phi}_\alpha]=& \tilde{S} (\rho_1^2 + \rho_2^2) +\frac{V_1}{4} (\rho_1^4 + \rho_2^4) + V_2 \rho_1^2 \rho_2^2 \\
	&+ W \rho_1^2 \rho_2^2 \cos(2 (\Theta_1 - \Theta_2)).
\end{align*}
\noindent Due to the last term, $V_{\tilde{\Phi}_\alpha}[\tilde{\Phi}_\alpha]$ is minimized when \be &&\Theta_1 = \Theta_2 {\rm~ or~} \Theta_1=\Theta_2 + \pi  {\rm ~~for~} W<0\nn && \Theta_1 = \Theta_2 \pm \frac{\pi}{2}{\rm ~~~~~~~~~~~~~~~~~for~}W>0.\label{potm}\ee Setting $\Theta_1$ and $\Theta_2$ to satisfy \Eq{potm} we obtain, for either sign of $W$, 
\begin{align*}
	V_{\tilde{\Phi}_\alpha}[\tilde{\Phi}_\alpha]= \tilde{S} (\rho_1^2 + \rho_2^2) +\frac{V_1}{4} (\rho_1^4 + \rho_2^4) + V_2 \rho_1^2 \rho_2^2 - |W| \rho_1^2 \rho_2^2. 
\end{align*}
\noindent It is straightforward to show the extrema of the above equation occur at
\begin{align*}
	\rho_1^2 = \rho_2^2 = \frac{-\tilde{S}}{\frac{V_1}{2} + V_2 - |W|} \text{ or } 0.
\end{align*}
\noindent The non-zero solution would require $$\tilde{S}\left(\frac{V_1}{2} + V_2 - |W|\right)<0.$$  Moreover, for the $V_{\tilde{\Phi}_\alpha}[\tilde{\Phi}_\alpha]$ to be bounded from below we must have $\frac{V_1}{2} + V_2 - |W|>0$. Therefore
\begin{align}
	\text{For $\tilde{S}>0$ : } & ~\tilde{\Phi}_1 = \tilde{\Phi}_2 = 0 \notag\\
	\text{For $\tilde{S}<0$ : } & 
	\begin{cases}
		\tilde{\Phi}_1 = \pm \tilde{\Phi}_2 := \rho e^{i\Theta}  &\text{ for } W<0 \\
		\tilde{\Phi}_1 = \pm i \tilde{\Phi}_2 := \rho e^{i\Theta}  &\text{ for } W>0 \\
	\end{cases}
	\label{Phi_SB}
\end{align}
\noindent Note that $\tilde{\Phi}_1 = \pm \tilde{\Phi}_2$ preserves time-reversal but breaks $90$ degree rotation symmetry, while $\tilde{\Phi}_1 = \pm i\tilde{\Phi}_2$ breaks time-reversal but preserves $90$ degree rotation symmetry. \\

\subsection{The condensation of p-wave double vortices -- a nematic, symmetry-enriched, $\mathbb{Z}_2$ spin liquid}\label{NNSL}
\hfill

The sign of $W$ relevant to this subsection is $W<0$. Let us consider the case where $\tilde{S}<0$ (this is the mean-field value, in reality there will be fluctuation correction). Let's take, e.g., 
\begin{align*}
	\tilde{\Phi}_1 = \tilde{\Phi}_2 = \rho e^{i\Theta}.
\end{align*}
Plug this into \Eq{Sphi-Phit} we obtain
\be
	\label{Sphi-Phit1}
	&&S_{\phi,\tilde{\Phi}} = \int d^3 x \Big\{ \frac{1}{2g} (\epsilon_{\mu\nu\rho} \partial_\nu a_\rho )^2+|(\partial_\mu - i a_\mu ) \phi_\sigma|^2\nn&& +2\rho^2|(\partial_\mu\Theta - 2 a_\mu ) |^2+V_{\phi_\s}[\phi_\s]\nn
	&&-\lambda\rho\left[ e^{-i\Theta} \left(\frac{1}{\sqrt{2}} \phi^T E \partial_x \phi \right) + \text{c.c.}~\right]
	\Big\}.
\ee
In passing to the last line we have used the fact that $\p_1+\p_2=\p_x$. \\

It is important to note that under inversion, $
\Theta\ra\Theta+\pi$. Since $\Theta$ can be absorbed by $a_\mu$, \Eq{Sphi-Phit1} is invariant under inversion. Therefore, although naively $p$-wave paired double vortex breaks the inversion symmetry, only the 90 degree rotation is broken because the sign change can be absorbed into the phase of $\tilde{\Phi}_{1,2}$. The resulting phase is nematic.\\

Isolating the parts of \Eq{Sphi-Phit1} that depends on $\phi_\sigma$ we obtain
\begin{align}
	& \int d^3 x \Big\{  |(\partial_\mu-i a_\mu) \phi_\sigma|^2 + V_{\phi_\s}[\phi_\s] \nn
	&-\lambda \rho \left[e^{-i\Theta}\left(\frac{1}{\sqrt{2}} \phi^T E \partial_1 \phi + \frac{1}{\sqrt{2}} \phi^T E \partial_2 \phi \right) +c.c\right]
	\Big\}\notag\\
	=& \int d^3 x \Big\{  |(\partial_\mu - i a_\mu)  \phi_\sigma|^2+ V_{\phi_\s}[\phi_\s]-\frac{1}{\sqrt{2}}\lambda \rho \Big[ e^{-i\Theta}\phi^T E \partial_x \phi \notag\\&+\text{c.c.}\Big]
	\Big\}.
	\label{phi_nematic}
\end{align}
In passing to the last line we have used the fact that $\p_1+\p_2=\p_x$.

Next, we perform the gauge transformation
\begin{align*}
	\phi_\s \rightarrow e^{i \frac{\Theta}{2}} \phi_\s,
\end{align*}
\noindent under which the action becomes
\begin{align}
	\label{phi-nematic1}
	S_\phi =& \int d^3x \Big\{ |(\partial_\mu - i A_\mu) \phi_\sigma |^2 + V_{\phi_\s}[\phi_\s]\notag\\
	&- \lambda  \sum_{i=1}^2 \left[ \rho \left( \frac{1}{\sqrt{2}} \phi^T E \partial_x \phi \right) + \text{c.c.} \right] \Big\} . 
\end{align}
\noindent where
\begin{align*}
	A_\mu := a_\mu - \frac{1}{2} \partial_\mu \Theta.
\end{align*} 
Note that  $\phi_\s$ is a spin 1/2 field under the global spin $SU(2)$ transformation. 
Moreover, because of the double vortex condensation, the gauge charge (with respect to $a_\mu$) of $\phi_\s$ becomes conserved $mod$ 2, i.e., equal to either 0 or 1.  \\

Let's now consider the case where $\phi_\s$ remains massive and ask what is momentum location where the mass gap is minimum. For this purpose, we temporarily turn off the coupling to $A_\mu$  in \Eq{phi-nematic1}.  Decomposing $\phi_\sigma$ into the real and imaginary parts
\be
\phi = \begin{pmatrix}  \phi_1 \\ \phi_2 \end{pmatrix} = \begin{pmatrix} u_1 + i u_3 \\ u_2 + i u_4 \end{pmatrix},
\label{realu}
\ee
\noindent the action that is quadratic in $u$ can be written as
\be
\int d^3x  \Big\{ -u^T \left( \partial^2 + s  II \right)u - \sqrt{2} \lambda \rho ~ u^T ZE \partial_x  u \Big\},
\label{ueq}
\ee
where $$u^T=(u_1,u_2,u_3,u_4).$$

\noindent Upon Fourier transformation \Eq{ueq} turns into
$$\int {d^3q\over (2\pi)^3}~u^T_{-q}~h(q)~u_{q}$$ where
\begin{align*}
	h(q_0,q_x,q_y) =& \left( q^2 + s \right) II + \sqrt{2} \lambda \rho ~ ZY q_x.
\end{align*}
\noindent The eigenvalues of $h(q_0,q_x,q_y)$ are
\begin{align*}
	&q_0^2+q_x^2 + q_y^2 + s \pm \sqrt{2} \lambda \rho q_x \\
	=& \left(q_x  \pm  \frac{1}{\sqrt{2}}\lambda \rho\right)^2 + q_0^2+q_y^2 +\left[s- \frac{1}{2}(\lambda \rho)^2\right]. 
\end{align*}

For $s>	\frac{1}{2} (\lambda \rho)^2$, $\phi_\sigma$ is gapped, and the minimum of the gap occurs at $\pm \v q_0$ where
\be \v q_0=(\frac{1}{\sqrt{2}}\lambda \rho,0).\label{qloc}\ee 
Now let's turn the coupling to $A_\mu$ back on, and integrate out the massive $\phi_\s$ to yield an effective theory for $A_\mu$. (For details see appendix \ref{boson_int}.)
\noindent To the leading order (in $\lambda\rho/\sqrt{s}$) the result is 
\be
	{\sqrt{s}\over 24\pi} \left(\frac{\lambda \rho}{\sqrt{s}}\right)^2 \int d^3x A_y^2={\sqrt{s}\over 96\pi} \left(\frac{\lambda \rho}{\sqrt{s}}\right)^2 \int \left( \partial_y \Theta - 2 a_y \right)^2\nonumber 
\ee
This spatial anisotropic Higgs term should be added to the 
\begin{align*}
	&\int d^3x |(\partial_\mu -  2 a_\mu ) \tilde{\Phi}_1|^2+|(\partial_\mu -  2 a_\mu ) \tilde{\Phi}_2|^2 \\
	=&2\rho^2\int d^3 x (\partial_\mu \Theta-  2 a_\mu )^2 
\end{align*}
term in \Eq{Sphi-Phit} to yield
\be
	&&\int d^3x\Big\{\left(2\rho^2\right)(\partial_\mu \Theta-  2 a_\mu )^2 +{\sqrt{s}\over 96\pi} \left(\frac{\lambda \rho}{\sqrt{s}}\right)^2 (\partial_y \Theta- 2 a_y)^2\Big\}\nn
	&&= \int d^3x ~~ \frac{g_\mu}{2} (\partial_\mu \Theta-2 a_\mu)^2
\label{nHiggs}\ee
\noindent 
where $\mu$ is summed over $0,1,2$.\\ 

Now we perform the standard duality transformation \cite{Fisher1989} on the Boltzmann weight associated with \Eq{nHiggs}. The first step includes a Hubbard-Stratonovich transformation  followed by decomposing $ e^{i \Theta}$ into a smooth and vortex part, namely, 
$e^{i \Theta}=e^{i \Theta_s}\times \Upsilon_v.$ This leads to 
\be
	&&\exp \Big\{ -\int d^3x ~~ \frac{g_\mu}{2} (\partial_\mu \Theta-2 a_\mu)^2\Big\}\nn
&&=\int D[J^\mu] \exp \Big\{ -\int d^3 x \Big[\frac{1}{g_\mu} (J^\mu)^2 
- i J^\mu \Big( \partial_\mu \Theta_s\nn&& + \frac{1}{i} \Upsilon_v^\dagger \frac{\partial_\mu}{i}  \Upsilon_v - 2 a_\mu \Big) \Big]\Big\}.
\ee
In the second step, we perform the integration over $\Theta_s$ and write the path integral over $\Upsilon_v$ as the sum over  the vortex world-lines (vw)
\begin{align}
	&\sum_{\rm vw}\int D[\Theta_s] D[J^\mu] ~\exp\Big\{-\int  \Big[  \frac{1}{2 g_\mu} (J^\mu)^2\notag\\
	& - i J^\mu \left( \p_\mu\Theta_{s} + \frac{1}{i} \Upsilon_v^\dagger{\partial_\mu\over i}\Upsilon_v - 2 a_\mu \right)  \Big] \Big\} \notag\\
	&=\sum_{\rm vw}\int\limits_{\partial_\mu J^\mu =0} D[J^\mu] ~ \exp\Big\{-\int  \Big[ \frac{1}{2g_\mu} (J^\mu)^2 \notag\\
	&  - i J^\mu \left(  \frac{1}{i} \Upsilon_v^\dagger{\partial_\mu\over i}\Upsilon_v- 2 a_\mu \right) \Big\}
	\label{inm1} 
\end{align}
Third, we solve the constraint 	$\partial_\mu J^\mu =0$ by introducing a gauge field
$$	J^\mu = \frac{1}{2\pi} \epsilon^{\mu\nu\rho} \partial_\nu b_\rho,$$
\Eq{inm1} becomes
\begin{align}
	&\sum_{\rm vw} \int D[b_\mu ]~ \exp\Big\{-\int  \Big[ \frac{1}{8\pi^2 g_\mu} ( \epsilon^{\mu\nu\rho} \partial_\nu b_\rho)^2 \notag\\
	&-  \frac{i}{2\pi} \epsilon^{\mu\nu\rho} \partial_\nu b_\rho \left( \frac{1}{i} \Upsilon_v^\dagger{\partial_\mu\over i}\Upsilon_v - 2 a_\mu \right)\Big] \Big\} \notag\\
	=&\sum_{\rm vw} \int D[b_\mu]~\exp\Big\{-\int  \Big[  \frac{1}{8\pi^2 g_\mu} ( \epsilon^{\mu\nu\rho} \partial_\nu b_\rho)^2  \notag\\ &+\frac{i}{\pi} \epsilon^{\mu\nu\rho} b_\mu \partial_\nu a_\rho + i  b_\mu  K^\mu_{\Upsilon_v}   \Big] \Big\} 
	\label{dualact}
\end{align}
where $K^\mu_{\Upsilon_v}$ is defined by
\begin{align*}
	K^\mu_{\Upsilon_v}=& \frac{1}{2\pi} \epsilon^{\mu\nu\rho}\partial_\nu \left( \frac{1}{i} \Upsilon_v^\dagger{\partial_\mu\over i}\Upsilon_v\right) .
\end{align*}
Physically $K^\mu_{\Upsilon_v}$ is the vortex current associated with  $\Upsilon_v$. To summarize, after including the Maxwell term for $a_\mu$ in \Eq{Sphi-Phit}, the total effective action read
\begin{align}
	\label{effa22}
	&S_{\rm eff}=\int d^3 x \Big[  \frac{1}{2g} (\epsilon_{\mu\nu\rho} \partial_\nu a_\rho )^2+\frac{1}{8\pi^2 g_\mu} ( \epsilon^{\mu\nu\rho} \partial_\nu b_\rho)^2 \\
	&+\frac{i}{\pi} \epsilon^{\mu\nu\rho} b_\mu \partial_\nu a_\rho + i  b_\mu  K^\mu_{\Upsilon_v}  \Big] . \notag
\end{align}
\Eq{effa22} can be rewritten as 
\be
	&&S_{\rm eff}=\int d^3 x \Big[  \frac{1}{2g} (\epsilon_{\mu\nu\rho} \partial_\nu a_\rho )^2+\frac{1}{8\pi^2 g_\mu} ( \epsilon^{\mu\nu\rho} \partial_\nu b_\rho)^2 \nn
	&&+{i\over 4\pi} \int d^3 x \Big[\epsilon_{\mu\nu\lambda}\sum_{I,J} a^I_\mu K_{IJ} \p_\nu a^J_\lambda+i  \sum_{\alpha}~j^\mu_\alpha  ({l}_{\alpha,I} a^I_\mu)\Big].\nn 	\label{effa222}
\ee
 Here  $I,J=1,2$ with $a^1_\mu=a_\mu$ and $a^2_\mu=b_\mu$, $\alpha=\Upsilon_v, \phi_\s$. In addition, 
\be
&&	l_{\phi_\sigma} = \begin{pmatrix} 1 \\ 0 \end{pmatrix},~~~
	l_{ K_{\Upsilon_v}} = \begin{pmatrix} 0 \\ 1 \end{pmatrix},~~~\nn
	&&K = \begin{pmatrix} 0 & 2 \\ 2 & 0 \end{pmatrix}  \Rightarrow K^{-1} = \begin{pmatrix} 0 & 1/2 \\ 1/2 & 0 \end{pmatrix}\nn&& j_\alpha^{\mu}(\v x,\t)=\delta^2(\v x-\v x_\alpha(\t)) (1, \dot{\v x}_\alpha).
\label{kmx}\ee In \Eq{kmx} $\v x_\alpha(\t)$ is the world line of the $\alpha$-type quasiparticle\footnote{For the sake of the statistics, we only care about how $\phi_\s$ couples to $a_\mu$, hence the current we write down for $j_{\phi_\s}$ is only the ``charge'' current with respect to $a_\mu$. To freeze the spin degrees of freedom one can polarize the spin in, say, the $\hat{z}$ direction.}. It is important to point out that in \Eq{effa222} we have put back the current of a  $\phi_\s$ {\it test-particle} in order to keep track of the statistics. At low energies and long wavelengths the first two Maxwell terms in \Eq{effa222} are irrelevant, hence can be omitted. The remaining parts of \Eq{effa222} determines the topological order. \Eq{effa222} is consistent with the matter-double-Chern-Simons action of Ref.\cite{Xu2009}.\\

This $K$ matrix in \Eq{kmx} contains the information of the ground state degeneracy on Riemann surfaces and the self and mutual statistics of quasiparticles \cite{Wen1992}. On a genus $g$ Riemann surface, the degeneracy $=|det(K)|^g =4^g$. 
The exchange statistics of the $\alpha$-type quasiparticle is given by the Berry phase $$\pi~ (l^T_\alpha \cdot K^{-1}\cdot l_\alpha).$$ The mutual statistics is given by the Berry phase 
$$2\pi ~(l_\alpha^T \cdot K^{-1}\cdot l_\beta)$$
arising from  particle of type $\alpha$ circling around particle of type $\beta$. 
From the above formula, it's easy to check that both particles are bosons. Circulating $\phi_\sigma$ around $K_{\Upsilon_v}$ gets a phase of $e^{i \pi} = -1$. This last fact implies that the bound state of $\phi_\sigma$ and $K_{\Upsilon_v}$ is a fermion, with
\begin{align*}
	l_{\phi_\sigma\otimes K_{\Upsilon_v} }= \begin{pmatrix} 1 \\ 1 \end{pmatrix}.
\end{align*}
Topologically, the statistics of the $\phi_\s$ particle, the $\Upsilon_v$ vortex, and the bound state of $\phi_\s$ and $\Upsilon_v$ vortex are the same as  the anyons $(e,m,\epsilon)$ in a $Z_2$ spin-liquid, namely, 
\be
&\phi_\sigma	\text{~particle}				\rightarrow ~ e \nn
&\Upsilon \text{~vortex}				\rightarrow ~ m \nn
&\text{Bound state of~}\Upsilon_v \text{~vortex and~} \phi_\s \text{~particle}\rightarrow ~ \epsilon=e\cdot m.\nn
\label{topo}
\ee

It is important to point out that in addition to the  $\mathbb{Z}_2$ topological order in \Eq{topo}, this spin liquid breaks spatial rotation symmetry as shown in \Eq{qloc} and \Eq{nHiggs}. Moreover, since $\phi_\s$ carries spin 1/2 quantum number, this anyon is symmetry-enriched. Therefore the spin liquid we obtain is a nematic, symmetry-enriched, $\mathbb{Z}_2$ spin liquid.\\

\subsection{The subsequent condensation of single vortex -- an uni-directional spiral AF phase}\label{icsdw}
\hfill

The subsequent condensation of $\phi_\s$ in the nematic symmetry-enriched $\mathbb{Z}_2$ spin liquid occurs when $s<\frac{1}{2} (\lambda \rho)^2$ (mean-field value, in reality there will be fluctuation correction). Under such condition the momentum modes $q_\mu= \pm\frac{1}{\sqrt{2}} \lambda \rho~ (0,1,0)$ of $\phi_\s$  condenses. Plugging into $\vec{n}=\phi^{\dagger}\vec{\s}\phi$ and taking into account of \Eq{prjt} and \Eq{qloc}, this gives incommensurate spiral with wave vector 
$$(\pi,\pi)\pm 2 \v q_0.$$  The phase diagram  is shown in \Fig{NSL}, which also appears in Ref.\cite{Sachdev1991, Xu2009, Thomson2018}\\

\subsection{The condensation of $p_x+ip_y$ double vortices -- a time-reversal breaking symmetry-enriched $\mathbb{Z}_2$ spin liquid }\label{TRBSL}
\hfill

The sign of $W$ in \Eq{Phi_SB} relevant to this subsection is  $W>0$. Let's consider $\tilde{S}<0$ (mean-field value, in reality there should be fluctuation correction) and choose, e.g., 
\begin{align*}
	\tilde{\Phi}_1 = -i\tilde{\Phi}_2 = \rho ~ e^{i\Theta}.
\end{align*}
\noindent Here we consider the situation that $\phi_\s$ remains massive. We determine the momentum location where the mass gap of $\phi_\s$ is minimized by turning off the coupling to $A_\mu$   in
\begin{align}
	\label{phiT1}
	S_\phi =& \int d^3x \Big\{ |(\partial_\mu - i A_\mu) \phi_\sigma |^2 + s|\phi_\sigma|^2 + \frac{v}{2} |\phi_\sigma|^4  \notag\\
	&- \lambda  \sum_{i=1}^2 \left[ \rho \left( \frac{1}{\sqrt{2}} \phi^T E (\partial_x+i\partial_y) \phi \right) + \text{c.c.} \right] \Big\} . 
\end{align}
After the decomposition in \Eq{realu}, the action that is quadratic in $u$ reads
\be
&\int d^3x  \Big\{ -u^T \left( \partial^2 + s  II \right)u  \notag\\
&- \sqrt{2} \lambda \rho ~ u^T \left( ZE ~ \partial_1 + XE ~ \partial_2 \right) u \Big\}.
\label{ueq2}
\ee
\noindent Upon Fourier transformation, \Eq{ueq2} turns into
$$\int {d^3q\over (2\pi)^3}~u^T_{-q}~h(q)~u_{q}$$ where
\be
	h(q_0,q_1,q_2) =& \left( q^2 + s \right) II + \sqrt{2} \lambda \rho ~ (ZY  ~ q_1 + XY ~ q_2).\nn\label{hqqq}
\ee
The eigenvalues of 	$h(q_0,q_1,q_2)$ are 
\begin{align*}
	&(q_0^2+|\v q|^2) + s \pm \sqrt{2} \lambda \rho ~ |\v q| \\
	=& q_0^2+\left(|\v q| \pm \frac{1}{\sqrt{2}} \lambda \rho\right)^2 +\left[s - \frac{1}{2} (\lambda \rho)^2\right]. 
\end{align*}
For $s> \frac{1}{2} (\lambda \rho)^2$,  the minimum gap of $\phi_\sigma$ occurs at 
$$|\v q|=\frac{1}{\sqrt{2}} \lambda \rho,$$ which is a ring in momentum space. \\

Because the eigenvector of \Eq{hqqq} is not invariant under the time reversal transformation \be h(q_0,\v q)\ra (ZE)^{-1}\cdot h(q_0,-\v q)\cdot ZE\label{Teg},\ee the $\phi_\s$ mode at momentum $\v q$ and spin $\s$ does not have the same energy as that at momentum $-\v q$ and spin $-\s$, i.e.,
\be
E_\s(\v q)\ne E_{-\s}(-\v q).
\label{trbk}
\ee
\Eq{trbk} is a manifestation of the breaking of time reversal symmetry.\\

Since $\phi_\s$ is gapped we can integrate it out 
\noindent to yield an effective action for $A_\mu$. To the leading order in $\frac{\rho\lambda}{\sqrt{s}}$ the answer is (the details are given in appendix \ref{boson_int})

\begin{align*}
	&{\sqrt{s}\over 24\pi} \left(\frac{\lambda \rho}{\sqrt{s}}\right)^2  \int d^3 x \left( A_1^2  + A_2^2\right)\\
	&= {\sqrt{s}\over 96\pi} \left(\frac{\lambda \rho}{\sqrt{s}}\right)^2 \int d^3 x \sum\limits_{i=1}^2\left( \partial_i \Theta - 2 a_i \right)^2. \\
\end{align*}

This Higgs term should be added to 
\begin{align*}
	&\int d^3 x\left\{|(\partial_\mu -  2 a_\mu ) \tilde{\Phi}_1|^2+|(\partial_\mu -  2 a_\mu ) \tilde{\Phi}_2|^2\right\}\\
	=&2\rho^2\int d^3 x (\partial_\mu \Theta-  2 a_\mu )^2 
\end{align*}
of \Eq{Sphi-Phit} to yield
\be
	&&\int d^3x\Big\{\left(2\rho^2\right)(\partial_\mu \Theta-  2 a_\mu )^2\nn&&+{\sqrt{s}\over 96\pi}~\frac{\rho^2\lambda^2}{s}\left[\sum_{i=1}^2 (\partial_i \Theta_i -2a_i)^2\right]\Big\}
\label{THiggs}\ee

\noindent This time, the additional contribution to the Higgs term breaks the space-time rotation but preserves the spatial rotational symmetry. In the subsequent duality transformation, all the derivations in subsection \ref{NNSL} follow, except that the Maxwell term for $b_\mu$ becomes space direction-independent. The K-matrix which determines the self and mutual statistics of quasiparticles remains unchanged.
Here the spin liquid is a time-reversal breaking, symmetry-enriched, $\mathbb{Z}_2$ spin liquid.

\subsection{The subsequent condensation of single vortex -- a ``double spiral'' AF phase}\label{dspsdw}
\hfill

\noindent For $s< \frac{1}{2}(\lambda \rho)^2$ (mean-field value, in reality there should be fluctuation correction), a whole ring of momentum modes of $\phi_s$, with $q_0=0$ and $|\v q| = \frac{1}{\sqrt{2}} \lambda \rho $,  become unstable. This preserves the spatial rotational symmetry but breaks the time-reversal symmetry (\Eq{trbk}). 
We have checked that up to the fourth-order terms in $u$, all the modes lying on the momentum ring remain degenerate. Plugging into $\vec{n}=\phi^{\dagger}\vec{\s}\phi$, this leads to a ring of momentum center around $(\pi,\pi)$. Following Ref.\cite{Kane1990} we refer to this phase as the ``double spiral'' AF phase. The phase diagram corresponds to $W>0$ is given in \Fig{TSL}.\\

\section{The spin triplet double vortex}
\label{AF*}
\hfill

Here we consider the scenario that the double vortex is a spin-triplet, even-parity, bound state of single vortices. In terms of  $\phi_\s$, the double vortex field is given by
\be
\Phi_n (\v x)= \frac{1}{\sqrt{2}} \phi_a^T (\v x)(E \sigma_n)_{ab}\phi_b(\v x),~~n={\rm x,y,z}.
\label{dbv}
\ee
\noindent Under the action of the symmetries in \Eq{sym}, $\Phi_n$ transforms according to
\be
&&\text{Vorticity conservation U(1): } \Phi_n \rightarrow  e^{i 2\eta} \Phi_n \nn
&&\text{Flavor $SU(2)$: }\Phi_m \rightarrow  R_{mn} \Phi_{n},~ R_{mn} \in \text{spin-1 rep of SU(2)} \nn
&&\text{$T_{\hat{x}}$, $T_{\hat{y}}$: }\Phi_n \rightarrow -\Phi_n^*\nn  
&&\text{$T$: }\Phi_n \rightarrow -\Phi_n.
\label{sym2}
\ee
 The action involving $\phi_\s$ and $\Phi_{n}$, and is invariant under \Eq{sym} and \Eq{sym2}, is given by
\begin{align}
\label{Sphi-Phi}
&S_{\phi,\Phi} = \int d^3 x \Big\{|(\partial_\mu - i a_\mu ) \phi_\sigma|^2+|(\partial_\mu - i 2 a_\mu ) \Phi_n|^2 \\
& + s |\phi_\sigma |^2 + \frac{v}{2} |\phi_\sigma|^4
+ S |\Phi_n |^2 \notag\\
&+ \frac{V_1}{2} (\Phi_n^* \Phi_n)(\Phi_m^* \Phi_m) + \frac{V_2}{2} (\Phi_n^* \Phi_n^*) (\Phi_m \Phi_m) \notag\\
&-\lambda \left[ \Phi_n^* \left(\frac{1}{\sqrt{2}} \phi_a^T (E \sigma_n)_{ab} \phi_{b} \right) + \text{h.c.}~\right] + \frac{1}{2g} (\epsilon_{\mu\nu\rho} \partial_\nu a_\rho )^2 \Big\} \notag
\end{align}
where the $\lambda$ term  enforces \Eq{dbv}. In the following, we consider the scenario where the double-vortex $\Phi_n$ condenses while the single-vortex $\phi_\sigma$ remain gapped.

\subsection{The condensation of spin triplet double vortices -- the AF* phase}
\label{K_matrix}
\hfill

For $S<0$ (mean-field value, in reality there should be fluctuation correction), the magnitude of $\Phi_n$ acquires a non-zero expectation value and the fluctuation of  $|\Phi_n|$ is massive.  Under such condition it is convenient to separate the massive and massless degrees of freedom by writing 
\be
\label{Phi_coherent}
&&\Phi_n = \rho_\Phi ~e^{i\Theta}  Z_n(\hat{\Omega}),\label{su2ph}\ee 
\noindent Here $Z(\hat{\Omega})$ is the spin-1 coherent state, which is related to the spin 1/2 coherent state, 
\begin{align*} 
	&z(\hat{\Omega}):= \begin{pmatrix} \cos(\theta/2) \\ \sin(\theta/2) e^{i\gamma}   \end{pmatrix}:= \begin{pmatrix} \alpha \\ \beta   \end{pmatrix}  \\
	 &\hat{\Omega}=(\sin \theta \cos \gamma, \sin \theta \sin \gamma, \cos\theta) 
\end{align*}
by
\be Z_{n}(\hat{\Omega})= \frac{1}{\sqrt{2}}z(\hat{\Omega})^T E \sigma_n z(\hat{\Omega})= \begin{pmatrix} \frac{1}{\sqrt{2}} (\alpha^2- \beta^2) \\ \frac{i}{\sqrt{2}} (\alpha^2 + \beta^2) \\ - \sqrt{2} \alpha \beta \end{pmatrix}.\label{Zz}\ee 

Plugging \Eq{Phi_coherent} and \Eq{Zz} into the stiffness terms for $\Phi_n$ in \Eq{Sphi-Phi}, we get
\be
\label{Higgs_term}
&&\int d^3 x |(\partial_\mu - i 2a_\mu) \Phi_n|^2 \nn
&&=\rho_\Phi^2 \int d^3 x\Big\{ \Big[
	 \left(\Upsilon^\dagger{\partial_\mu\over i}\Upsilon- 2 a_\mu +  \frac{1}{i} Z^\dagger \partial_\mu Z\right)^2\nn
&&+(\partial_\mu Z^\dagger) (\partial_\mu Z) + (Z^\dagger \partial_\mu Z) (Z^\dagger \partial_\mu Z)
\Big] \Big\} \nn
&&=\rho_\Phi^2 \int d^3 x\Big\{\Big[ 
	\Big(\Upsilon^\dagger{\partial_\mu\over i}\Upsilon  - 2 a_\mu +  \frac{1}{i} Z^\dagger \partial_\mu Z\Big)^2 
	\nn&&+ \frac{1}{2} (\partial_\mu \hat{\Omega})^2 \Big] \Big\},
\ee
where $$\Upsilon:=e^{i\Theta}.$$
Note that
\begin{align*}
 \left( \frac{1}{i} Z^\dagger dZ \right)= 2\left(\frac{1}{i} z^\dagger dz\right) = 2 \sin^2 \frac{\theta}{2} ~ d\gamma
\end{align*}
\\

In the phase where $\phi_\s$ is massive, we can integrate out $\phi_\s$. The parts of action involving the single vortex field $\phi_\sigma$ read
\begin{align*}
	&S_\phi = \int d^3x \Big\{ |(\partial_\mu - i a_\mu) \phi_\sigma |^2 + s|\phi_\sigma|^2 + \frac{v}{2} |\phi_\sigma|^4 \\
	&- \lambda \rho_\Phi \left[ e^{-i \Theta} Z_n^* \left( \frac{1}{\sqrt{2}} \phi^T E \sigma_n \phi \right) + h.c. \right] \Big\}
\end{align*}	

To leading order (in $\lambda\rho_\Phi/s$), the result of $\phi_\sigma$ integration (see appendix \ref{boson_int} for details) is
\begin{align*}
    &-{\sqrt{s}\over 32 \pi}\left(\frac{\lambda \rho_\Phi}{s}\right)^2 \int d^3 x \Big[ 7 (\partial_\mu \hat{\Omega} )^2 \\
    &+ 3 \left(\Upsilon^\dagger{\partial_\mu\over i}\Upsilon  - 2 a_\mu +  \frac{1}{i} Z^\dagger \partial_\mu Z\right)^2\Big]
\end{align*}
Adding the above result to \Eq{Higgs_term}, we have
\begin{align}
\label{effa}
\int d^3 x\Big[	&\frac{1}{2\lambda_\Theta}\left(\Upsilon^\dagger{\partial_\mu\over i}\Upsilon  - 2 a_\mu +  \frac{1}{i} Z^\dagger \partial_\mu Z\right)^2 \\
	&+ \frac{1}{2 \lambda_{\Omega}} (\partial_\mu \hat{\Omega})^2 \Big] \notag
\end{align}
\noindent where 
\begin{align*}
{1\over 2\lambda_\Theta}=&\rho_\Phi^2 \left( 1 - \frac{3 \lambda^2}{32 \pi s^{3/2}} \right) \\
{1\over 2\lambda_\Omega}=& \frac{\rho_\Phi^2}{2} \left( 1 -  \frac{7 \lambda^2}{16 \pi s^{3/2}}\right).
\end{align*}

Now  we generalize the duality transformation for $S=1/2$  bosons in Ref.\cite{Lee1990} to $S=1$ bosons. First we  introduce the Hubbard-Stratonavich field $J^\mu$ and  split 
$$\Upsilon=e^{i \Theta_s}\Upsilon_v
$$ 
to obtain 
\begin{align*}
&\exp\Big\{-\int d^3 x ~  \frac{1}{2 \lambda_{\Omega}} (\partial_\mu \hat{\Omega})^2 \notag\\
&+\frac{1}{2\lambda_\Theta} \left( \frac{1}{i}\Upsilon^\dagger{\partial_\mu\over i}\Upsilon - 2 a_\mu + \frac{2}{i} z^\dagger \partial_\mu z \right)^2  \Big\} \notag\\
&=\int D[J^\mu] ~\exp\Big\{-\int d^3 x  ~\Big[   \frac{1}{2 \lambda_{\Omega}} (\partial_\mu \hat{\Omega})^2 \notag\\
&\frac{\lambda_\Theta}{2} (J^\mu)^2- i J^\mu \left( \p_\mu\Theta_s + \frac{1}{i} \Upsilon_v^\dagger{\partial_\mu\over i}\Upsilon_v- 2 a_\mu + \frac{2}{i} z^\dagger \partial_\mu z \right) \Big] \Big\}.
\end{align*}
Second, we perform the integration over $\Theta_s$ and write the path integral over $\Upsilon_v$ as the sum over the vortex world lines,
\be
&&\sum_{\rm vw}\int D[\Theta_s] D[J^\mu] ~\exp\Big\{-\int  \Big[ \frac{1}{2 \lambda_{\Omega}} (\partial_\mu \hat{\Omega})^2+\frac{\lambda_\Theta}{2} (J^\mu)^2  \nn
&&- i J^\mu \left( \p_\mu\Theta_{s} + \frac{1}{i} \Upsilon_v^\dagger{\partial_\mu\over i}\Upsilon_v - 2 a_\mu + \frac{2}{i} z^\dagger \partial_\mu z \right)  \Big] \Big\}\nn
&&=\sum_{\rm vw}\int\limits_{\partial_\mu J^\mu =0} D[J^\mu] ~ \exp\Big\{-\int  \Big[+\frac{1}{2 \lambda_{\Omega}} (\partial_\mu \hat{\Omega})^2\Big] \Big\} \nn
&&+ \frac{\lambda_\Theta}{2} (J^\mu)^2 - i J^\mu \left(  \frac{1}{i} \Upsilon_v^\dagger{\partial_\mu\over i}\Upsilon_v- 2 a_\mu + \frac{2}{i} z^\dagger \partial_\mu z\right) 
\label{inm} 
\ee
Third, we solve the constraint 	$\partial_\mu J^\mu =0$ by introducing a gauge field
$$	J^\mu = \frac{1}{2\pi} \epsilon^{\mu\nu\rho} \partial_\nu b_\rho,$$
\noindent after which \Eq{inm} becomes
\begin{align}
&\sum_{\rm vw} \int D[b_\mu ]~ \exp\Big\{-\int  \Big[ \frac{1}{2 \lambda_{\Omega}} (\partial_\mu \hat{\Omega})^2 +\frac{\lambda_\Theta}{8\pi^2} ( \epsilon^{\mu\nu\rho} \partial_\nu b_\rho)^2 \notag\\
&-  \frac{i}{2\pi} \epsilon^{\mu\nu\rho} \partial_\nu b_\rho \left( \frac{1}{i} \Upsilon_v^\dagger{\partial_\mu\over i}\Upsilon_v - 2 a_\mu + \frac{2}{i} z^\dagger \partial_\mu z\right)\Big] \Big\} \notag\\
=&\sum_{\rm vw} \int D[b_\mu] ~\exp\Big\{-\int  \Big[ \frac{1}{2 \lambda_{\Omega}} (\partial_\mu \hat{\Omega})^2 + \frac{\lambda_\Theta}{8\pi^2} ( \epsilon^{\mu\nu\rho} \partial_\nu b_\rho)^2 \notag\\
&- \frac{i}{2\pi} \epsilon^{\mu\nu\rho} b_\mu \partial_\nu\left( \frac{1}{i} \Upsilon_v^\dagger{\partial_\mu\over i}\Upsilon_v - 2 a_\rho +\frac{2}{i} z^\dagger \partial_\rho z \right) \Big] \Big\} \nn
=&\sum_{\rm vw} \int D[b_\mu]~\exp\Big\{-\int  \Big[ +\frac{1}{2 \lambda_{\Omega}} (\partial_\mu \hat{\Omega})^2+\frac{\lambda_\Theta}{8\pi^2} ( \epsilon^{\mu\nu\rho} \partial_\nu b_\rho)^2 \notag\\ &+\frac{i}{\pi} \epsilon^{\mu\nu\rho} b_\mu \partial_\nu a_\rho + i  b_\mu \left( K^\mu_{\Upsilon_v} + 2  K^\mu_{\hat{\Omega}} \right) \Big] \Big\} 
\label{dualact}
\end{align}
where $K^\mu_{\Upsilon_v}$ and $K^\mu_{\hat{\Omega}}$ are defined by
\begin{align*}
&K^\mu_{\Upsilon_v}= \frac{1}{2\pi} \epsilon^{\mu\nu\rho}\partial_\nu \left( \frac{1}{i} \Upsilon_v^\dagger{\partial_\mu\over i}\Upsilon_v\right) \\
&K^\mu_{\hat{\Omega}} =  \frac{1}{2\pi} \epsilon^{\mu\nu\rho}\partial_\nu\left(\frac{1}{i} z^\dagger \partial_\rho z\right) = \frac{1}{4\pi} \epsilon^{\mu\nu\rho}  \sin\theta ~ \partial_\nu \theta ~ \partial_\rho \gamma \\
&= \frac{1}{8\pi} \epsilon^{\mu\nu\rho} \epsilon^{abc} \hat{\Omega}^a. ~ \partial_\nu \hat{\Omega^b} ~ \partial_\rho \hat{\Omega^c}
\end{align*}
Physically $K^\mu_{\Upsilon_v}$ is the vortex current in $\Upsilon_v$, and $K^\mu_{\hat{\Omega}}$ is the skyrmion current in $\hat{\Omega}$. To summarize, after including the Maxwell term for $a_\mu$ in \Eq{Sphi-Phi}, the total effective action reads
\begin{align}
	\label{effa2}
&S_{\rm eff}=\int d^3 x \Big[  \frac{1}{2g} (\epsilon_{\mu\nu\rho} \partial_\nu a_\rho )^2+\frac{\lambda_\Theta}{8\pi^2} ( \epsilon^{\mu\nu\rho} \partial_\nu b_\rho)^2 \\
&+\frac{i}{\pi} \epsilon^{\mu\nu\rho} b_\mu \partial_\nu a_\rho + i  b_\mu \left( K^\mu_{\Upsilon_v} + 2  K^\mu_{\hat{\Omega}} \right) +\frac{1}{2 \lambda_{\Omega}} (\partial_\mu \hat{\Omega})^2\Big] . \notag
\end{align}

The parts  of \Eq{effa2} that determine the topological order is the same as \Eq{kmx} except now there are three quasiparticles, namely, $\alpha=\phi_\s$, vortex in $\Upsilon_v$ and skyrmion in  $\hat{\Omega} $.  The current of these quasiparticles couples to the gauge fields via \Eq{kmx}
where
\begin{align*}
l_{\phi_\sigma} = \begin{pmatrix} 1 \\ 0 \end{pmatrix},~~~
l_{ K_{\Upsilon_v}} = \begin{pmatrix} 0 \\ 1 \end{pmatrix},~~~
l_{K_\Omega} = \begin{pmatrix} 0 \\ 2 \end{pmatrix}.
\end{align*}
Follow the same procedure as in subsection \ref{NNSL}, it's easy to check that all these three quasiparticle are bosons. $K_\Omega$ has mutual boson statistics with all other particles, while the mutual Berry phase $\phi_\sigma$ between and $K_{\Upsilon_v}$ is $e^{i \pi} = -1$. This last fact implies that the bound state of $\phi_\sigma$ and $K_{\Upsilon_v}$  with
\begin{align*}
l_{\phi_\sigma\otimes K_{\Upsilon_v} }= \begin{pmatrix} 1 \\ 1 \end{pmatrix},
\end{align*}
is a fermion.
Topologically, the self/mutual statistics of the $\phi_\s$ particle, the $\Upsilon_v$ vortex, and the bound state of $\phi_\s$ and $\Upsilon_v$ vortex are the same as in the anyon quasiparticles $(e,m,\epsilon)$ in a $Z_2$ spin-liquid, namely, 
\be
&&\hat{\Omega} \text{~skyrmion} 		\rightarrow ~ 1 \nn
&&\phi_\sigma	\text{~particle}				\rightarrow ~ e \nn
&&\Upsilon_v \text{~vortex}				\rightarrow ~ m \nn
&&\text{Bound state of~}\Upsilon_v \text{~vortex and~} \phi_\s \text{~particle}\rightarrow ~ \epsilon=e\cdot m.\nn
\ee

In this spin liquid there is a  coexisting N\'{e}el long-range order. The Goldstone mode of such order is described by the term proportional to $\int d^3x (\p_\mu\hat{\Omega})^2$ in \Eq{effa2}.  This coexisting 
symmetry-enriched $\mathbb{Z}_2$ topological order and N\'{e}el order is referred to as ``AF*'' in the literature.

\subsection{The subsequent condensation of single vortex -- back to the N\'{e}el state}\label{btn}
\hfill

From the AF* phase discussed in the preceding subsection, the condensation of $\phi_\s$ will Higgs the gauge field $a_\mu$, and consequently the $\mathbb{Z}_2$ topological order is lost. The resulting phase is the same as the N\'{e}el state obtained directly from condensing the single VBS vortex (see \Fig{AFSL}). The reason there is no more phase transition in the lower half of the phase diagram in \Fig{AFSL} is the following. Once the single vortex $\phi_\s$ condenses, the $\lambda$ term in \Eq{Sphi-Phi} acts as an "magnetic field term" for $\Phi_n$. Hence $\Phi_n$ will automatically acquire a non-zero expectation value. Consequently, there is no transition when one tunes $S$ from positive to negative.  \\

As mentioned earlier, the scenario depicted in \Fig{AFSL} requires two $S=1/2$  vortices on the same sublattice (hence the same vorticity) to form a triplet bound state. It is unclear to us what kind of spin interaction will favor that.

\section{Final Discussion}\label{condis}

	In this section, we will not talk about the AF* phase in section \ref{AF*} due to the irrelevancy to frustrated antiferromagnets. \\

As mentioned earlier, it is quite remarkable that the result of subsection \ref{NNSL} and \ref{icsdw} agree with the result of Schwinger boson parton approach in Ref.\cite{Sachdev1991}.  It turns out that the spin-1/2 VBS vortex in the current work plays the role of the Schwinger boson in Ref.\cite{Sachdev1991}.  The $U(1)$ gauge field, which is dual to the phase of the VBS order parameter in the current work, plays the role of the gauge field arising from the fractionalization of the physical, spin, into bilinear in Schwinger bosons.
Since the spin liquid state possesses $\pi$-flux anyon, the bosonic spinon can form a bound state with it and become a fermion. The quantum number of such excitation is consistent with the fermionic spinon obtained in the fermionic parton approach of  $\mathbb{Z}_2$ spin liquid \cite{Wen1991,Thomson2018}. 
\\

The nematicity in the spin liquid discussed in subsection \ref{NNSL} is due to a spontaneous symmetry breaking. It is possible for quantum fluctuations to suppress the nematic order and restore the 90-degree rotation symmetry. In that case, the spin liquid discussed in subsection \ref{NNSL} will become a true symmetry-enriched $\mathbb{Z}_2$ spin liquid (i.e., without  symmetry breaking). Similarly, quantum fluctuations can also restore the time-reversal symmetry and render the spin liquid discussed in subsection \ref{TRBSL} a true symmetry-enriched $\mathbb{Z}_2$ spin liquid.   \\

Finally, we would like to say a few words about hole doping. Motivated by the results of subsection \ref{fermion_zero_mode}, we expect doping to remove the $S=1/2$ in the VBS vortex core. Moreover, we can show that after such spin removal, the vortex becomes a spinless fermion. Since the vorticity remains unchanged by such a removal process, naively one expects these fermions to inherit the vorticity quantum number, namely, $f_+$ and $f_-$, and live on opposite sublattices. However, in the spin liquid phase, the double vortex has condensed. This makes the vorticity only defined  $mod$ 2. As a result, $f_+$ and $f_-$ are no longer distinct. The resulting single fermion species  live on both sublattices. This is consistent with the 
bosonic spinon approach where the doped holes are spinless fermions. When such a  fermion bounds with a neutral $S=1/2$ bosonic vortex, the vorticity cancels. (Note that in the presence of double vortex condensation, the vorticity of a single vortex is also defined $mod$ 2.) 
The composite particle has charge $+e$ and spin 1/2, hence has the same quantum number as the photohole in photoemission experiments. Because of the vorticity cancellation, to the leading order, these fermionic holes no longer couple to the VBS gauge field (which has been Higgsed to $\mathbb{Z}_2$). These composite holes can form a topological Fermi liquids state (FL*) featuring a small Fermi surface \cite{Punk2015}. 

In the N\'{e}el, the spiral, and the double spiral SDW phases, where the single vortex condenses, the distinction between $f_+$ and $f_-$ also ceases to exist.  When such a  fermion bounds with the condensed $S=1/2$ single vortex, it also acquires the quantum number of the photohole. These composite holes carry spin 1/2 moment and interact with the order parameter $\vec{n}=\phi^\dagger\vec{\s}\phi$ via the Yukawa coupling. In those phases, the geometry of the small Fermi surfaces will depend on the magnetic order. In the VBS ordered phase the two types of fermion $f_+$ and $f_-$ remain distinct, and they hop among the same sublattice. Such a fermion can also form a composite hole with a $S=1/2$ vortex residing on the opposite sublattice (to cancel the vorticity). It is unclear to us whether such composite holes can be itinerant in the presence of VBS order. If so, they will form a conventional FL with small Fermi surfaces due to broken symmetry. \\

In the spin liquid phases, the $\pi$-flux anyon can convert the spinless fermion into boson, and at the same time convert $\phi_\s$ into $S=1/2$ fermions. These quantum numbers are consistent with the fermionic spinon approach, where the  doped holes are bosons. \\

\section*{Acknowledgement}

\noindent  This work was primarily funded by the U.S. Department of Energy, Office of Science, Office of Basic Energy Sciences, Materials Sciences and Engineering Division under Contract No. DE-AC02-05-CH11231 (Theory of Materials program KC2301). This research is also funded in part by the Gordon and Betty Moore Foundation.\\

\begin{appendices}

\section{Integrating out the massive $\phi_\s$}
\label{boson_int}

\subsection{The general formalism}

In integrating out $\phi_\s$, we face the following functional integral
\begin{align}
&Z[a_\mu, \Theta]=e^{-W[a_\mu, \Theta]} = \int Du(x)~e^{-S[u(x),a_\mu, \Theta]} {\rm ~~where~~ }\notag\\
&S[u(x),a_\mu, \Theta] = \int d^3 x \, u^T~\mathcal{D}[a_\mu, \Theta]~u.
\label{ReEff}
\end{align}
Here we have adapted \Eq{realu} so that  $u(x)$ is a real boson field and $\mathcal{D}[a_\mu, \Theta]$ is a symmetric  space-time (differential) operator. A convenient trick for doing such integration is to perform the corresponding complex boson integration and divide the resulting effective action by two. \\

To see this, consider two copies of real boson fields $u_\a$ and $u_\b$ coupled identically to  $a_\mu$ and $\Theta$. After the boson integration, the result should be the square of that in \Eq{ReEff}, namely,
\begin{align*}
	&\int Du_\a \, Du_\b \, e^{-\left[ S[u_\a(x),a_\mu, \Theta] + S[u_\b(x),a_\mu, \Theta]\right]} \\
	&=\left\{ Z[a_\mu, \Theta] \right\}^2=e^{-2W[a_\mu, \Theta]}
\end{align*}
On the other hand, we can combine $u_{\a,\b}$ into a complex fermion field $$\varphi= u_\a + i u_\b,$$ so that the sum of the real fermion actions can be written as a complex fermion action,
\begin{align*}
	&u_\a^T \mathcal{D}[a_\mu, \Theta] u_\a + u_\b^T \mathcal{D}[a_\mu, \Theta] u_\b\\
	&=\varphi^\dagger\mathcal{D}[a_\mu, \Theta]\varphi.  
\end{align*}
Note that the cross terms cancel out, due to the commutation relation between $u_\a$ and $u_\b$, and the fact that $$\Big[\mathcal{D}[a_\mu, \Theta]\Big]^T=\mathcal{D}[a_\mu, \Theta].$$ 
Consequently if  $\tilde{W}[a_\mu, \Theta]$ is the effective action due to the complex boson integration, we have
\begin{align}
&W[a_\mu, \Theta]  = \frac{1}{2} \tilde{W}[a_\mu, \Theta] \notag\\
&= -\frac{1}{2}\ln \Big[ \det(\mathcal{D}[a_\mu, \Theta])^{-1}  \Big] = \frac{1}{2} \Tr\Big[\ln  (\mathcal{D}[a_\mu, \Theta])  \Big] .
\label{crr}\end{align}
Due to \Eq{crr}, we shall focus on the complex boson integration.\\

To proceed we write 
\be
		\mathcal{D}[a_\mu, \Theta] = \mathcal{D}_0 + V[a_\mu, \Theta] = (-\partial^2 + s) + V[a_\mu, \Theta]
\label{defv}\ee
\noindent where $V[a_\mu, \Theta]$ includes the terms describing the coupling of $\phi_\s$ to the gauge field and $\Theta$. By the usual perturbative expansion we have
\begin{align*}
	W[a_\mu, \Theta]  =&\frac{1}{2} \Tr\Big[\ln  (\mathcal{D}_0 + V[a_\mu, \Theta])  \Big]  \\
	=& \frac{1}{2} \Tr \Big[ \ln(\mathcal{D}_0) + \sum\limits_{n=1}^\infty \frac{(-1)^{n+1}}{n} \left( \mathcal{D}_0^{-1} V\right)^n \Big].
\end{align*}
The control parameter of this expansion will be discussed separately for the spin-triplet parity-even and spin-singlet odd parity double vortex cases. Finally, the $\Tr\Big[ \ln(\mathcal{D}_0)\Big]$ term is a constant independent of the external field so that we can neglect it. 

\subsection{Spin-singlet odd-parity double vortex condensation}

For spin-triplet-parity-even double vortex the part of action that involves $\phi_\sigma$ is given by 
\begin{align*}
	S_\phi =& \int d^3x \Big\{ |(\partial_\mu - i a_\mu) \phi_\sigma |^2 + s|\phi_\sigma|^2  \\
	&- \lambda  \sum_{i=1}^2 \left[ \rho_i e^{-\Theta_\alpha} \left( \frac{1}{\sqrt{2}} \phi^T E \partial_i \phi \right) + h.c. \right] \Big\}. 
\end{align*}
\noindent Here we do the transformation
\begin{align*}
	\phi \rightarrow e^{i \frac{\Theta}{2}} \phi,
\end{align*}
\noindent under which the action becomes
\begin{align*}
	S_\phi =& \int d^3x \Big\{ |(\partial_\mu - i A_\mu) \phi_\sigma |^2 + s|\phi_\sigma|^2  \\
	&- \lambda  \sum_{i=1}^2 \left[ \rho_i  \left( \frac{1}{\sqrt{2}} \phi^T E \partial_i \phi \right) + h.c. \right] \Big\} 
\end{align*}
\noindent where
\be
	A_\mu := a_\mu - \frac{1}{2} \partial_\mu \Theta
	\label{defv2}
\ee is a U(1) gauge field.\\

In terms of the real boson field in \Eq{ReEff}, the potential $V[a_\mu,\Theta]$ in \Eq{defv} is 
\begin{align*}
	V[a_\mu,\Theta] =& \int d^3 x\Big[-i\sqrt{2} \lambda  \sum_{i=1}^2 \rho_{\alpha i } \tilde{M}_{\alpha} \partial_i \\
	&+ i(\partial_\mu A_\mu + A_\mu \partial_\mu) \Sigma_0 + A_\mu A_\nu\Big]  
\end{align*}
where $A_\mu$ is given by \Eq{defv2},
\noindent and  $\rho_{1 i}$ and $\rho_{2 i}$ are the real and imaginary parts of $\rho_i$, and
\begin{align*}
	&\Sigma_0 = -YI \\
	&\tilde{M}_1= ZY \\
	&\tilde{M}_2= XY
\end{align*}
\noindent The remaining real boson integration is straightforward. The leading order (in $\lambda\rho_i/\sqrt{s}$) contribution gives
\begin{align*}
	W=&{\sqrt{s}\over 24\pi} \frac{\lambda^2}{s} \int A_i \left[ \rho_{\alpha l} \rho_{\alpha l} \delta_{ij} - \rho_{\alpha i} \rho_{\alpha j} \right] A_j 
\end{align*}\\

For the nematic spin liquid in \Eq{phi_nematic}, we have $\rho_i = \delta_{ix} \rho$. This gives us
\begin{align*}
	W=&{\sqrt{s}\over 24\pi} \left(\frac{\lambda \rho}{\sqrt{s}}\right)^2 \int d^3 x A_y^2 \\
	=&{\sqrt{s}\over 96\pi} \left(\frac{\lambda \rho}{\sqrt{s}}\right)^2 \int d^3 x\left( \partial_y \Theta - 2 a_y \right)^2 
\end{align*}
In the case of the time-reversal breaking spin liquid we have $\rho_1=\rho$ and $\rho_2=i\rho$. This results in
\begin{align*}
	W=& {\sqrt{s}\over 24\pi} \left(\frac{\lambda \rho}{\sqrt{s}}\right)^2  \int d^3 x \left( A_1^2  + A_2^2\right) \\
	=&{\sqrt{s}\over 96\pi} \left(\frac{\lambda \rho}{\sqrt{s}}\right)^2 \int d^3 x \sum\limits_{i=1}^2\left( \partial_i \Theta - 2 a_i \right)^2. 
\end{align*}
Again the presence of the prefactor $\sqrt{s}$ can be determined from dimension counting.
\\

\subsection{Spin-triplet even-parity double vortex condensation}

For spin-triplet-parity-even double vortex, the term in the action which involves   $\phi_\sigma$ is given by 
\begin{align*}
	S_\phi =& \int d^3x \Big\{ |(\partial_\mu - i a_\mu) \phi_\sigma |^2 + s|\phi_\sigma|^2  \\
	&- \lambda \rho_\Phi \left[ e^{-i \Theta} Z_n^*(\hat{\Omega}) \left( \frac{1}{\sqrt{2}} \phi^T E \sigma_n \phi \right) + h.c. \right] \Big\} 
\end{align*}
\noindent Before continuing, note that we can do the transformation
\begin{align*}
	\phi \rightarrow e^{i \frac{\Theta}{2}} u^\dagger \phi,
\end{align*}
\noindent where
\begin{align*}
	u(\hat{\Omega}) = \begin{pmatrix} 
			\cos \frac{\theta}{2} 				& -\sin \frac{\theta}{2} e^{-i \phi} \\
			\sin \frac{\theta}{2} e^{i \phi}	& \cos \frac{\theta}{2}
	 \end{pmatrix}
\end{align*}
\noindent and action becomes
\begin{align*}
	S_\phi =& \int d^3x \Big\{ |(\partial_\mu - i A_\mu) \phi_\sigma |^2 + s|\phi_\sigma|^2  \\
	&- \lambda \rho_\Phi \left[  Z_n^*(\hat{z}) \left( \frac{1}{\sqrt{2}} \phi^T E \sigma_n \phi \right) + h.c. \right] \Big\} 
\end{align*}
\noindent Here $$Z_n(\hat{z}) = \frac{1}{\sqrt{2}} (1, i, 0),$$ and $A_\mu$ contains both the $U(1)$ and the $SU(2)$ parts, namely,
\begin{align*}
	A_\mu :=& a_\mu - \frac{1}{2} \partial_\mu \Theta - \frac{1}{i} u \partial_\mu u^\dagger \\
	=& \sum\limits_{a=0}^{3}A_\mu^a \sigma_a
\end{align*}
\noindent where
\begin{align*}
	&A_\mu^0 = a_\mu - \frac{1}{2} \p_\mu\Theta \\
	&A_\mu^1= \frac{1}{2} \left( \sin\gamma ~ \partial_\mu  \theta + \sin\theta \cos\gamma ~ \partial_\mu \gamma\right) \\
	&A_\mu^2= \frac{1}{2} \left( -\cos\gamma ~ \partial_\mu \theta + \sin\theta \sin\gamma ~ \partial_\mu \gamma\right) \\
	&A_\mu^3= - \frac{1}{2} (1-\cos \theta) \partial_\mu \gamma = - \frac{1}{i} z^\dagger \partial_\mu z.	
\end{align*}\\

In terms of the real boson field in \ref{ReEff}, the coupling potential is
\begin{align*}
	V[a_\mu , \Theta, \hat{\Omega}] =& -\sqrt{2} \lambda  Z_{\alpha n}(\hat{z}) M_{\alpha n} \\
	&+ i(\partial_\mu A^a_\mu + A^a_\mu \partial_\mu) \Sigma_a + A_\mu^a A_\nu^b \Sigma_a \Sigma_b
\end{align*}
\noindent where $\alpha=1,2$ and $a,b,n=1,2,3$. The $Z_{1 n}(\hat{z})$ and $Z_{2 n}(\hat{z})$ are the real and imaginary parts of $Z_{n}(\hat{z})$, and
\begin{align*}
	&M_{1n} = (ZZ, -XI, -ZX) \\
	&M_{2n} = (-XZ, -ZI, XX) \\
	&\Sigma_a =  (-YI, -YX, IY, -YZ)
\end{align*}
\noindent The remaining perturbative integration of the boson is straightforward. The leading order (in $\lambda\rho_\Phi/s$) contribution gives
\begin{align*}
	&W[\Theta, \Omega, a_{\mu}]\\
	=&-{\sqrt{s}\over 32 \pi}\left(\frac{\lambda \rho_\Phi}{s}\right)^2\int \left[ 28 \left((A_\mu^1)^2 + (A_\mu^2)^2\right)  +12 (A_\mu^0 + A_\mu^3)^2 \right] \\
	=&-{\sqrt{s}\over 32 \pi}\left(\frac{\lambda \rho_\Phi}{s}\right)^2\int \left[ 7 (\partial_\mu \Omega)^2  +3 (2a_\mu - \partial_\mu \Theta - \frac{2}{i} z^\dagger \partial_\mu z)^2 \right] 
\end{align*}
The presence of the prefactor ${\sqrt{s}}$ is can be determined by dimension counting\footnote{The reason the dimensionless parameter is $\lambda\rho_i/s$ instead of $\lambda\rho_i/\sqrt{s}$ is  that in the absence of derivative, the dimension of the coupling $\lambda$ changes. }
\end{appendices}



\bibliographystyle{ieeetr}
\bibliography{bibs}

\end{document}